\documentclass[aps,prb,reprint,showpacs,floatfix]{revtex4-1}  % for PRB format
\usepackage{amsmath,amssymb,xcolor}
\usepackage[colorlinks,bookmarks=false,citecolor=red,linkcolor=blue,urlcolor=blue]{hyperref}
\usepackage{graphicx,epstopdf}
\usepackage{mathtools}
\usepackage{hyperref}
\hypersetup{
    colorlinks,
    citecolor=blue,
    filecolor=blue,
    linkcolor=blue,
    urlcolor=blue}
\begin{document}
\title{Orbital and spin ordering physics of the Mn$_3$O$_4$ spinel}
\author{Santanu Pal}
\email{sp13rs010@iiserkol.ac.in}
\author{Siddhartha Lal}
\email{slal@iiserkol.ac.in}
\affiliation{Department of Physical Sciences, Indian Institute of Science Education and Research-Kolkata, W.B. 741246, India}

%=============================================================
\begin{abstract}
Motivated by recent experiments, we present a comprehensive theoretical study of the geometrically frustrated strongly correlated magnetic insulator Mn$_3$O$_4$ spinel oxide based on a microscopic Hamiltonian involving lattice, spin and orbital degrees of freedom. Possessing the physics of degenerate e$_g$ orbitals, this system shows a strong Jahn-Teller effect at high temperatures. Further, careful attention is paid to the special nature of the superexchange physics arising from the 90$^o$ Mn-O-Mn bonding angle. The Jahn-Teller and superexchange-based orbital-spin Hamiltonians are then analysed in order to track the dynamics of orbital and spin ordering. We find that a high-temperature structural transition results in orbital ordering whose nature is mixed with respect to the two originally degenerate $e_{g}$ orbitals. This ordering of orbitals is shown to relieve the intrinsic geometric frustration of the spins on the spinel lattice, leading to ferrimagnetic Yafet-Kittel ordering at low-temperatures. Finally, we develop a model for a magnetoelastic coupling in Mn$_3$O$_4$, enabling a systematic understanding of the experimentally observed complexity in the low-temperature structural and magnetic phenomenology of this spinel. Our analysis predicts that a quantum fluctuation-driven orbital-spin liquid phase may be stabilised at low temperatures upon the application of pressure.
\end{abstract}
\maketitle

%\tableofcontents
%===========================================================================================
\section{Introduction}
Frustrated magnetic systems with orbital degeneracy present an interesting play ground for the exploration of novel ordered as well as liquid-like states. Frustration leads very naturally to macroscopic degeneracy in the ground state and a concomitant lack of equilibrium spin ordering. Such spin liquid states have been proposed in theoretical studies of several systems with different types of geometrically frustrated lattices, as well as in experiments performed on some candidate materials~\cite{balents2010spin,zhou2016quantum,fu2015evidence}. It is important to note, however, that typical material systems of interest in quantum magnetism also possess orbital and lattice degrees of freedom. The interaction among these three can lead to a variety of emergent ordered states. For instance, degenerate orbital degrees of freedom interact in a cooperative manner with fluctuations of the lattice via the Jahn-Teller effect~\cite{jahn1937stability,bersuker2006,fazekas1999lecture}, inducing orbital ordering along with a static global distortion of the lattice. From the seminal work of Kugel and Khomskii~\cite{kugel1973crystal, kugel1982jahn}, it is well known that strong electronic correlations give rise to superexchange-related interactions between and among the orbital and spin degrees of freedom. The resultant ordering of orbitals and spins can then be strongly tied to each other. Further, the coupling of spin and lattice degrees of freedom can also be shown to have interesting consequences for spin ordering~\cite{PhysRevLett.85.4960,tchernyshyov2011spin}. In this way, the presence of multiple couplings between various degrees of freedom leads generically to the separation of energy scales at which the ordering of the orbitals and spins takes place~\cite{oles2010charge,oles2012fingerprints,oles2000quantum,khomskii2001orbital}.
Equally importantly, such interactions also offer multiple ways by which the system can relieve any inherent frustration among the spins and attain the ordering of spins. 
\par
Since orbital-spin interactions depend strongly on the metal-ligand-metal bonding angle in strongly correlated insulators~\cite{mostovoy2002orbital,reitsma2005orbital}, a transition metal insulator on the geometrically frustrated spinel lattice with orbital, spin and lattice degrees of freedom offers the exciting prospect of finding diverse physical phenomena across a wide scale of energies. In this light, the spinel Mn$_3$O$_4$ is an ideal candidate system in which the complex interplay of various spin-orbital-lattice interactions have been studied~\cite{jensen1974magnetic,chardon1986mn3o4,suzuki2008magnetodielectric,kim2011pressure, chung2013low,nii2013interplay,byrum2016effects}. However, the microscopic mechanism that relieves the geometric frustration inherent in the spinel structure and leads to the ferrimagnetic Yafet-Kittel ordering of spins at low temperatures remain unknown~\cite{chartier1999ab}. Further, the superexchange mechanism for electronic correlations in the Mn$_3$O$_4$ spinel is likely different from the other well-studied transition-metal(TM) perovskite systems~\cite{feiner1997quantum,rosciszewski2010jahn}. The active orbitals on the nearest-neighbour TM sites in the former are orthogonally directed with respect to one another, while they are aligned in the latter. This renders inapplicable our intuition of orbital ordering based on the phenomenological Goodenough-Kanamori-Anderson(GKA) rules learnt from studies of perovskite systems.
\par 
Experiments show that at 1443K, Mn$_3$O$_4$ undergoes a structural transition from cubic to tetragonal lattice symmetry~\cite{chardon1986mn3o4,jensen1974magnetic,guillou2011magnetic,kim2011pressure, kim2010mapping, suzuki2008magnetodielectric, chung2013low,jo2011spin}. This material also has three different magnetic transitions, with the first being  a transition from a paramagnetic phase to a three dimensional canted ferrimagnetic Yafet-Kittel spin ordering at 43K. In the Yafet-Kittel phase, the magnetic unit cell possesses two Mn$^{2+}$(A-type) spins 
%at the tetrahedral site is 
aligned along the [110͔] direction, 
%in the cubic settings, and 
together with a tetrahedra of four Mn$^{3+}$(B-type) spins 
%at the octahedral site 
whose net moment is antiparallel to the Mn$^{2+}$ spins, but with each of the four spins being canted away from the c-axis ([001] direction) and towards the $[\bar{1}\bar{1}0͔]$ direction. This compound has  further magnetic transitions at 39K and 33K. Between these two transitions, the magnetic unit cell becomes incommensurate with respect to the chemical unit cell. Another structural transition, from tetragonal to orthorhombic lattice symmetry~\cite{kim2011pressure}, is observed at 33K and the Yafet-Kittel magnetic order is again attained, but with a unit cell doubled along the [110] direction with respect to the ordered phase at 43K. Experiments also show that the magnetic transitions are associated with a significant change in the dielectric constant, indicating a strong spin-lattice coupling~\cite{tackett2007magnetodielectric, suzuki2008magnetodielectric, chung2013low, nii2013interplay}.
\par
\begin{figure}[h!]
\centering
\includegraphics[scale=.3]{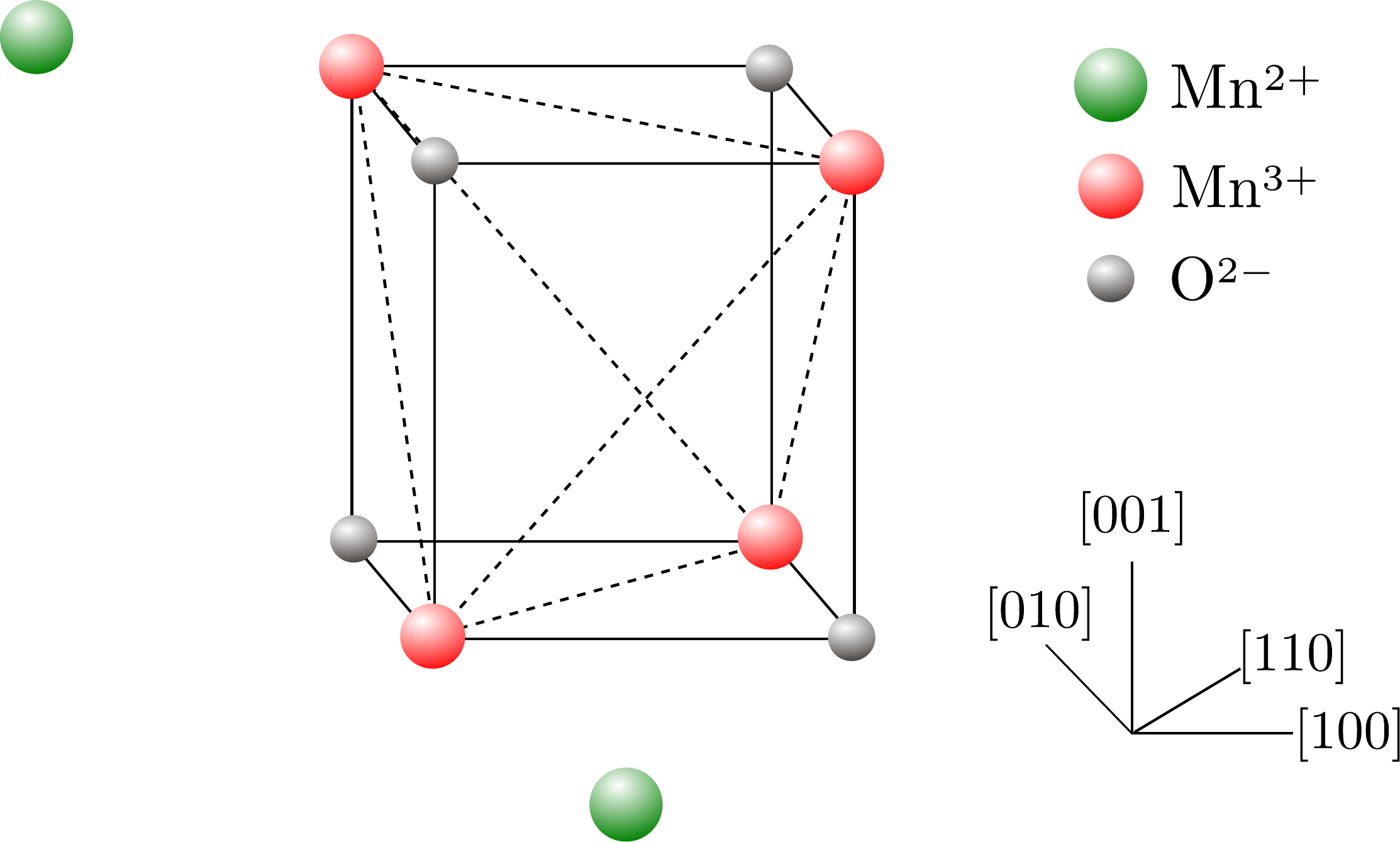}
\caption{ (Color online.) Schematic diagram of a Mn$_3$O$_4$ unit cell. Four orbitally active  Mn$^{3+}$ ions (red spheres) form a tetrahedron. The Mn$^{3+}$ ions are connected via Oxygen ions (gray spheres), with the Mn-O-Mn bonding angle being 90$^o$. There are also two orbitally non-active  Mn$^{2+}$ ions (green spheres) present in the unit cell.} 
\label{mnspinel}
\end{figure}

 
In this work, we attempt a qualitative understanding of underlying physics of ordering in the Mn$_3$O$_4$ spinel by the development of a microscopic model for the obrital, spin and lattice degrees of freedom. Keeping in mind the special 90$^o$ TM atom-ligand atom-TM atom bonding angle in Mn$_3$O$_4$ (a typical Mn$^{3+}$ tetrahedra is shown in Fig.(\ref{mnspinel})), we carry out a systematic derivation of the orbital-spin Hamiltonian. We then carry out a variational analysis of this Hamiltonian to find the nature of orbital ordering at higher energy scales and resultant spin ordering at lower energies. In understanding the low temperature structural transition (tetragonal to orthorhombic) and associated magnetic and orbital orders, we also develop a model for the spin-lattice interaction in this system. This paper is organised as follow. In section II, we derive the orbital-spin model. In section III, we analyse this Hamiltonian and discuss  our results for orbital ordering. Using the orbital order found in section III, we then compute the magnetic interaction in different crystallographic planes in section IV, along with a discussion of spin ordering at low temperatures. In section V, we develop a model for the spin-lattice interaction and relate our results to explain some experimental facts. We end with a concluding section.
%===========================================================================================
\section{\label{sec:spinorbitalHam} Spin-Orbital Hamiltonian}
\begin{figure}[!h]
 \centering
 \includegraphics[scale=.17]{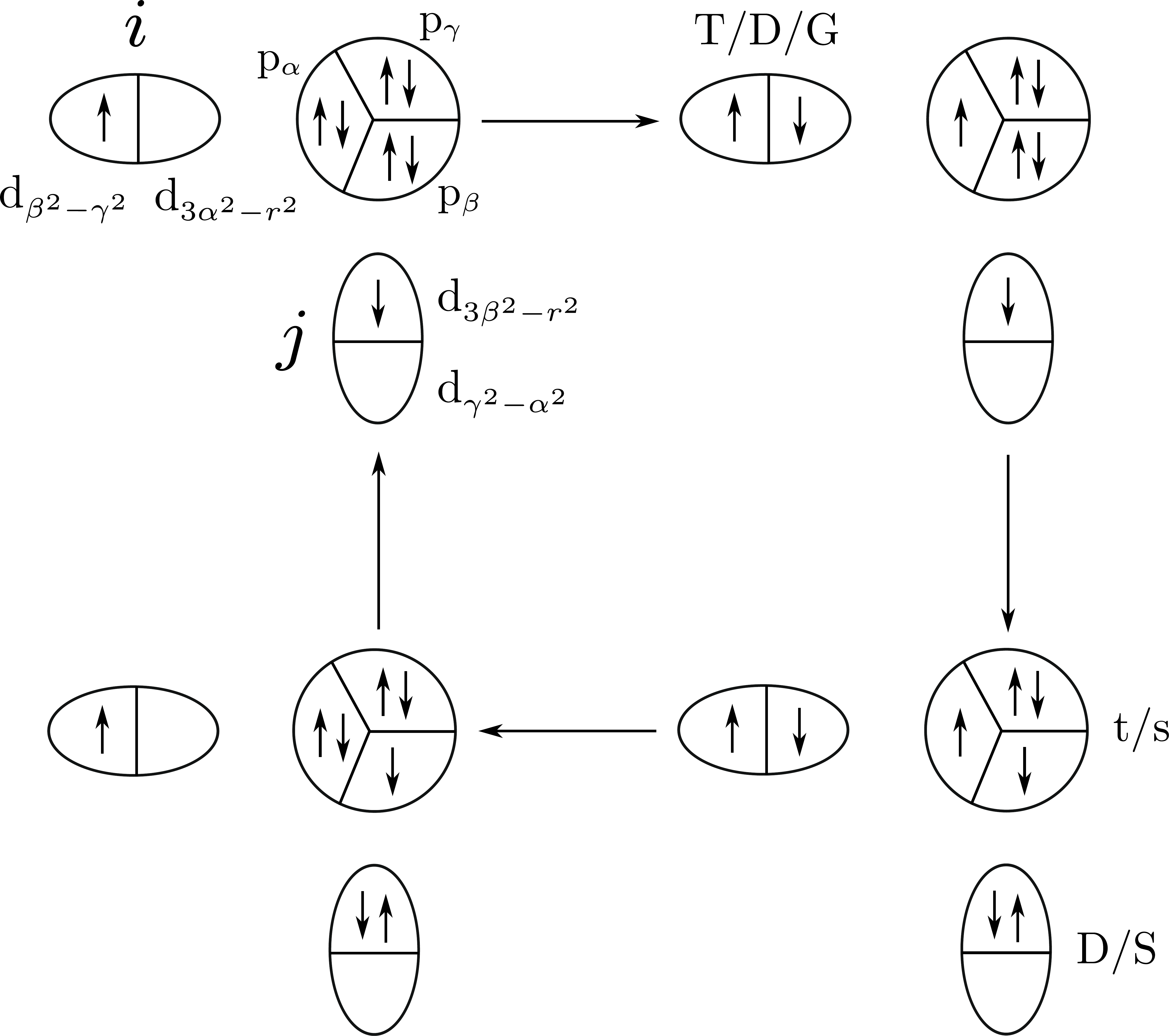}
 \caption{Four-step SE process for a M-type initial configuration (following notations of ~\cite{reitsma2005orbital}). An oval represents the Mn$^{3+}$ ion with one e$_g$ electron either in the non-hopping orbital (d$_{\beta^2-\gamma^2}$) on the left or the hopping orbital (d$_{3\alpha^2-r^2}$) on the right. A circle represents an Oxygen ion with the electrons in the $p$-orbitals. T, D, G and S are excited states of the Mn$^{3+}$ ion. The triplet $t$ and singlet $s$ are excited state configurations of the Oxygen ion. Note that $\{\alpha, \beta, \gamma\}$ a cyclic permutation of $\{ x, y ,z\}$.}
 \label{fig:fourstape}
 \end{figure}

The Mn$_3$O$_4$ spin-orbital model is based on the spinel lattice, where nearest neighbour transition metal ions are connected via Oxygen ligand atoms with ion-Oxygen-ion bond angle of 90$^o$. Due to the presence of a crystal field, the 3d electron levels of Mn$^{3+}$ ion split into t$_{2g}$ and e$_g$ levels \cite{nii2013interplay}. Further, for an on-site Hubbard repulsion on the transition metal ion that is much greater than the crystal field splitting, the Mn$^{3+}$ ion's 4 d-electrons form a high spin configuration with spin S$=$2~~\cite{feiner1999electronic}. As shown by Mostovoy \emph{et al.}~\cite{mostovoy2002orbital} and  Reitsma \emph{et al.} \cite{reitsma2005orbital} for the case of a 90$^o$ TM-O-TM bond angle, the Anderson superexchange process (where an electron is effectively transferred from one metal ion site to the other metal ion site) is considerably weaker than the Goodenough superexchange process (where two electron transfer from the ligand Oxygen ion, one to each of the neighbour metal ions). The Goodenough superexchange process is schematically represented as
%\begin{eqnarray}
\begin{equation}
e^1p^6e^1\rightarrow e^1p^5e^2\rightarrow e^2p^4e^2\rightarrow   e^1p^5e^2\rightarrow e^1p^6e^1~,
\label{eqn:Goodenoughprocess}
\end{equation}
where $e$ and $p$ represent the $e_{g}$ orbital on the transition metal (Mn$^{3+}$) and $p$ orbital on the ligand ($O$) sites respectively.
%\end{eqnarray} 
The effective spin-orbital Hamiltonian is obtained from fourth-order perturbation theory. As shown in figure (\ref{fig:fourstape}), in the excitation process of the superexchange mechanism (SE), one electron hops from a p$_{\alpha}$ orbital of the ligand Oxygen atom to the left Mn$^{3+}$ ion e$_g$ orbital, while another electron hops from a p$_\beta$ orbital of the same Oxygen atom to the down Mn$^{3+}$ ion orbital. In the de-excitation process, excited electrons return to the Oxygen atom  by reversing the hopping pathway. In this way, correlations build between the two Mn$^{3+}$ ions and contribute to the SE.
 The relevant  Mn$^{3+}$ e$_g$ states are $^4A_2,\, ^4E ,\, ^4A_1$ and $^6A_1$. Of these, the first is a singlet (S), the second and third are Hund's-split orbital doublets (D,G) and the fourth is a triplet (T), with energy $E_{S}= U+\frac{10}{3}J_H$, $E_{D}= U+\frac{2}{3}J_H$ ,$E_{G}=U$ and  $E_{T}=U-5J_H$~~\cite{feiner1999electronic, zaanen1990systematics} respectively. The oxygen p$^4$ configurations that occurr during the superexchange process include a  triplet $^3T_1$ (t) and singlet $^1T_2$ (s) with energy $U_p-J_p$ and $U_p+J_p$ respectively. Here $U$, $J_H$ and $U_p$, $J_p$ are the interorbital Coulomb repulsion and Hund's exchange couplings on the Mn$^{3+}$ and Oxygen ion sites respectively. 
In deriving the complete spin-orbital Hamiltonian, we follow the formulation and notations of Ref.(\cite{reitsma2005orbital}). While the detailed derivation is presented in an appendix, we present here the results. It should be noted that, in reaching the spin-orbital Hamiltonian, one must list all possible initial configurations (electron's position on hopping or non-hopping orbital of Mn-sites), and for each of them list all possible hopping sequences that return to the ground state manifold. Thus, the spin-orbital Hamiltonian we begin with has the form 
 \begin{eqnarray}
 \mathcal{H}_{eff}& =&\sum_{\langle ij\rangle}(\mathcal{Q}_{O,ij}^{\alpha\beta}[K_O^TQ_{ij}^T+K_O^SQ_{ij}^S]+\mathcal{Q}_{M,ij}^{\alpha\beta}[K_M^T Q_{ij}^T\nonumber \\
 &&+K_M^S Q_{ij}^S]+\mathcal{Q}_{N,ij}^{\alpha\beta}[K_N^T Q_{ij}^T+K_N^S Q_{ij}^S])
 \end{eqnarray}
where $\mathcal{Q}$'s are the orbital projection operator denoted by 
\begin{eqnarray}
 &&\mathcal{Q}_{O,ij}^{\alpha\beta}=\mathcal{P}_i^{\alpha} \mathcal{P}_j^{\beta} +  \mathcal{P}_i^{\beta} \mathcal{P}_j^{\alpha},\nonumber\\&& \mathcal{Q}_{M,ij}^{\alpha\beta}=\mathcal{P}_i^{\alpha} \mathcal{P}_j^{\bar{\beta}} +  \mathcal{P}_i^{\bar{\alpha}} \mathcal{P}_j^{\beta} +\mathcal{P}_i^{\beta} \mathcal{P}_j^{\bar{\alpha}} +  \mathcal{P}_i^{\bar{\beta}} \mathcal{P}_j^{\alpha},\nonumber\\ && \mathcal{Q}_{N,ij}^{\alpha\beta}=\mathcal{P}_i^{\bar{\alpha}} \mathcal{P}_j^{\bar{\beta}} +  \mathcal{P}_i^{\bar{\beta}} \mathcal{P}_j^{\bar{\alpha}}~,  
\end{eqnarray}
where $\mathcal{P}_i^\alpha=(\frac{1}{2}\textbf{I}_i+I_i^\alpha)$ and $\mathcal{P}_i^{\bar{\alpha}}=(\frac{1}{2}\textbf{I}_i-I_i^\alpha)$ are projection operators for the hopping and non-hopping orbitals respectively. The (rotated) orbital pseudospin operators ($I$) are given by
 \begin{align}
 I_i^{x/y}= -\frac{1}{2}T_i^z\mp\frac{\sqrt{3}}{2}T_i^x, \hspace*{.5cm} I_i^z=T_i^z~,
 \end{align}
 where $T$ represents the pseudospin operators for the two-fold degenerate e$_g$ orbital system, with the orbital Hilbert space at each site given by 
 \begin{align}
 \begin{pmatrix}
 1\\0
 \end{pmatrix}= |3z^2-r^2\rangle~~,\hspace*{.5cm}
 \begin{pmatrix}
 0\\1
 \end{pmatrix}= |x^2-y^2\rangle~~.
 \end{align}
 The orbital projection operators decide the precise configuration of the orbitals on any two Mn$^{3+}$ ions connected by a hopping pathway, e.g., whether both orbitals are occupied by an electron each (denoted by ``O"), any one of the two orbitals occupied (denoted by ``M") and neither orbital occupied (denoted by ``N"). 
%or one hopping and one non hopping orbital  or both electrons are at non-hopping orbital  and 
Further, $Q$ are spin projection operators denoted by
\begin{equation}
Q_{ij}^T=\frac{1	}{10}(6+\vec{S}_i.\vec{S}_j),\qquad Q_{ij}^S=\frac{1}{10}(4-\vec{S}_i.\vec{S}_j)~,
 \end{equation}
where $T$ stands for triplet and $S$ for singlet spin configurations at Mn-sites. Finally, the $K$s are various spin-orbital superexchange (SE) constants.
\par 
The Hamiltonian may be separated into purely-orbital and spin-orbital parts as follows
\begin{eqnarray}
\mathcal{H}_{eff}&=&\sum_{\langle ij\rangle}([J_O^o\mathcal{Q}_{O,ij}^{\alpha\beta}+J_M^o\mathcal{Q}_{M,ij}^{\alpha\beta}+
 J_N^o\mathcal{Q}_{N,ij}^{\alpha\beta}]\textbf{1}_{ij}\nonumber \\&&+[J_O^S\mathcal{Q}_{O,ij}^{\alpha\beta}+J_M^S\mathcal{Q}_{M,ij}^{\alpha\beta}+J_N^S\mathcal{Q}_{N,ij}^{\alpha\beta}]\vec{S}_i.\vec{S}_j),
 \label{eqn:spinorbitalHamiltonian}
\end{eqnarray}
with inter-orbital and spin-orbital interaction constants given by
\begin{eqnarray}
J_L^o=\frac{1}{10}(6K_L^T+ &&4K_L^S),\qquad J_L^S=\frac{1}{10}(K_L^T-K_L^S)\nonumber \\&& (L=O, M, N)~~.
 \label{eqn:SEconstant}
\end{eqnarray}
In order to calculate the SE constant, one  has to consider closely at the four-step superexchange process. As an example, consider the $M$ intial configuration (as shown in figure (\ref{fig:fourstape})). Here, a single electron (with spin-up) occupies a non-hopping orbital (d$_{\beta^2-\gamma^2}$) on the $i$th  site of Mn$^{3+}$, while  another electron (with spin-down) occupies the hopping orbital (d$_{3\beta^2-r^2}$) on the $j$th site; we refer to this initial state as $|M,\uparrow\downarrow\rangle$. The corresponding contribution to the SE coupling is denoted by $[M,\uparrow\downarrow;\downarrow\uparrow]$, where the final ($\downarrow\uparrow$) denote the configuration of hopping electrons from the oxygen sites. As we will see below, $[M,\uparrow\downarrow;\downarrow\uparrow]$ involve contributions from 12 possible doubly excited states denoted succinctly by $[XvY]$, with $X,Y \epsilon \{T,D,S,G\}$ and $v \epsilon \{t,s\}$~: 
 \begin{eqnarray}
 [M,\uparrow\downarrow;\downarrow\uparrow]&=&\frac{1}{3}([TtD]+[TtS]+[DtD]+[DtS]+[TsD]\nonumber \\ &&+[TsS]+[DsD]+[DsS]+[GtD]\nonumber \\ &&+[GtS]+[GsD]+[GsS])
 \end{eqnarray}
 with
 \begin{align}
[XvY] = \frac{t^4}{4}\frac{\Delta_X+\Delta_Y}{\Delta_X^2\Delta_Y^2}\frac{U_v}{\Delta_X+\Delta_Y+U_p-J_p}
 \end{align}
 where $U_s= U_p+J_p$, $U_t= U_p-J_p$ and  $\Delta_T=\Delta-4J_H,  \Delta_D=\Delta+\frac{5}{3}J_H, \Delta_S=\Delta+\frac{13}{3}J_H,  \Delta_G=\Delta+J_H$ and $\Delta=U-J_H$. Here $t$ is the charge transfer amplitude. The factor $1/3$ takes care of the three-fold processes arising from the $T$, $D$ and $G$ spin configurations of an excited state (see Fig.(\ref{fig:fourstape})). Note that we have four sequences of excitation and de-excitation processes for a given intermediate doubly-excited state. The ``O" and ``N" configurations are taken account of in a similar manner (see appendix (\ref{appendix}) for more detail).
 %Electron from oxygen ion can first excited to either ith site or to jth site, so two excitation sequence and de-excitation can also happens two different order. In general different order of sequence are energetically different, because first and third intermediate state may depend upon sequence. Taking care of this thing for all other configurations (O, N and other configuration of M) we have 
%  The four sequences that lead to a particular middle state upon inverting the order of the exciting and/or de-exciting charge transfers, are in general inequivalent, because the energies of the first and third intermediate state may depend upon the sequence. 
\par
The complete spin-orbital Hamiltonian can then be written in a compact notation as follows  
\begin{eqnarray}
\mathcal{H}_{eff}&=& \sum_{\langle i,j \rangle, \alpha\neq\beta} J_\tau \mathcal{W}_{ij}^{\alpha\beta}\nonumber \\ &&+\sum_{\langle ij \rangle,\alpha\neq\beta } (-J_\sigma\textbf{1}_{ij}+J_\nu\mathcal{V}_{ij}^{\alpha\beta}-J_\mu\mathcal{W}_{ij}^{\alpha\beta})\vec{S}_i.\vec{S}_j~,\nonumber \\
 \label{equn:hamiltonian}
\end{eqnarray}
where $\mathcal{V}^{\alpha\beta}_{ij}=-\textbf{1}_i(I_j^\alpha +I_j^\beta)-(I_i^\alpha+I_i^\beta)\textbf{1}_j$ and $\mathcal{W}^{\alpha\beta}_{ij}=2(I_i^\alpha I_j^\beta+I_i^\beta I_j^\alpha)$, with the first term denotes inter-orbital interactions and the second term to spin-orbital interactions. It is also worth noting that the form of the inter-orbital superexchange Hamiltonian and the Jahn-Teller Hamiltonian derived from phonon-orbital interactions for the spinel lattice~\cite{englman1970cooperative} are the same.
\par
\begin{figure}[h!]
 \centering
 \includegraphics[scale=.5]{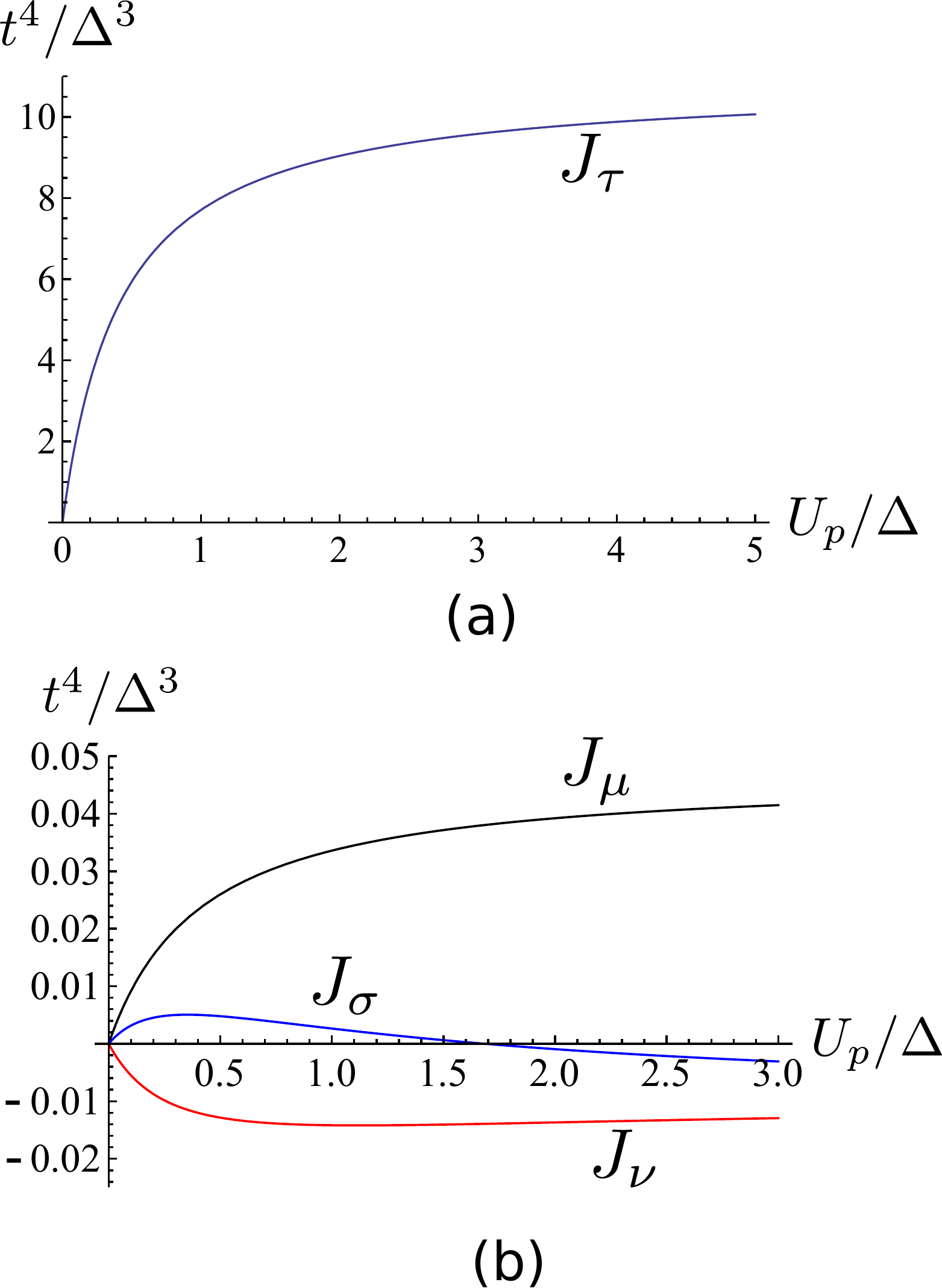}
 \caption{(Color online.) A plot of various superexchange interaction constants (in units of $J=t^4/\Delta^3$) as a function of $U_p/\Delta$  for $J_p/U_p=0.1$ and $J_H/\Delta=0.14$~: (a) Orbital superexchange constant $J_\tau$; (b) spin-orbital superexchange constants $J_\mu$, $J_\nu$ and  $J_\sigma$. Estimates of some of the parameters for Mn$_3$O$_4$ from the literature are $U_p/\Delta =0.909$, $J_H=0.69 eV$, $U=7.3 eV$, and $\Delta=5.5eV$.~\cite{bocquet-1992,mizokawa-1996,feiner1999electronic,reitsma2005orbital}}
  \label{fig:SEconstant}
 \end{figure}

  
The complete form of the various $J$ superexchange constants are presented in the appendix. It is, however, important to note that among all these SE couplings, only $J_\nu$ is negative for the Mn$_3$O$_4$ spinel (see figure (\ref{fig:SEconstant})).   
%$I^{\alpha/\beta}$ are rotated pseudospin operator for orbital degrees of freedom associated with the three cubic axes ($\alpha/\beta=$ a,b, or c),First part of the Hamiltonian is purely due to orbital-orbital interaction and one may call orbital only Hamiltonian. Last part is due to both spin and orbital  and may call spin-orbital Hamiltonian. 
Further, it is clear from figure (\ref{fig:SEconstant}) that the inter-orbital interaction $J_{\tau}$ is much greater than all spin-orbital interactions. Thus, orbital ordering must dominate the physics of this Hamiltonian at high energy-scales. 
% \begin{figure}[h!]
% \centering
%    \includegraphics[scale=.6]{seconstant1}
%\end{figure}
 %===========================================================================================
 \section{Variational Analysis of the inter-orbital Hamiltonian}
 Recalling the fact that among different superexchange constants, the inter-orbital superexchange term (J$_\tau$) dominates over all other exchange terms (see figure(\ref{fig:SEconstant})), we analyse the inter-orbital part of the Hamiltonian first in order to obtain the orbital ordering that will set in at high energy-scales. The form of the inter-orbital  mean field Hamiltonian is given by
 \begin{eqnarray}
 \mathcal{H}_{MF}= \sum_{\langle i,j \rangle, \alpha\neq\beta} J_\tau \langle\mathcal{W}_{ij}^{\alpha\beta}\rangle ~.
 \end{eqnarray}
We use a general superposition of the orbital basis-state for a variational analysis
 \begin{equation}
 |\theta_i\rangle = \cos(\theta_i/2)|3z^2-r^2\rangle + \sin(\theta_i/2)|x^2-y^2\rangle ~,
 \label{basis}
 \end{equation}
such that the expectation value of the orbital operator $\mathcal{W}_{ij}$ acting on the direct-product orbital state specifying the bond lying between the $i$th and $j$th sites, $|\theta_i\rangle \otimes |\theta_j\rangle$, is obtained as
 \begin{eqnarray}
 \langle \mathcal{W}_{ij}^{\alpha\beta}\rangle &&= 
\frac{1}{4}[2\cos(\theta_i+\theta_j)-\cos(\theta_i-\theta_j)]_{xy}\nonumber \\  &&+\frac{1}{4}[2\cos(\theta_i+\theta_j-\frac{4\pi}{3})-\cos(\theta_i-\theta_j)]_{xz}\nonumber \\ &&+\frac{1}{4}[2\cos(\theta_i+\theta_j+\frac{4\pi}{3})-\cos(\theta_i-\theta_j)]_{yz}~.
\label{eqn:expectation}
 \end{eqnarray}
Here, the suffix at the end of each bracket indicates the plane on which the particular Mn-O-Mn bond lies within a given Mn$^{3+}$ tetrahedron.
\par 
It can be shown that, for the case of 90$^0$ bonding between the transition metal ions, a ferro-orbital (FO) configuration ($\theta_i\simeq \theta_j$) is favoured energetically over a canted-orbital (CO) configuration ($\theta_i\simeq \theta_j\pm \pi$)\cite{kugel1982jahn, reitsma2005orbital}. Considering these subtleties in our analysis, we compute the variational ground state orbital ordering for the Mn$_3$O$_4$ spinel. The variational energy for this FO configuration of a tetragonally distorted spinel (lattice lengthscale along the $c$-axis is greater than that along the $a$ and $b$ axes, $c>a=b$) is found to be
\begin{align}
E(\theta)=\frac{J_\tau}{4}[2(1-\beta)\cos2\theta-(1+2\beta)] ~,
\end{align}
where the $XY$-plane superexchange constant is given by $J_\tau^{ab}=J_\tau$, and the $XZ$ and $YZ$-plane constants by $J_\tau^{ac}=J_\tau^{bc}=\beta J_\tau$~, $\beta<1$~.
\par
\begin{figure}[h!]
 \begin{center}
 \includegraphics[scale=.25]{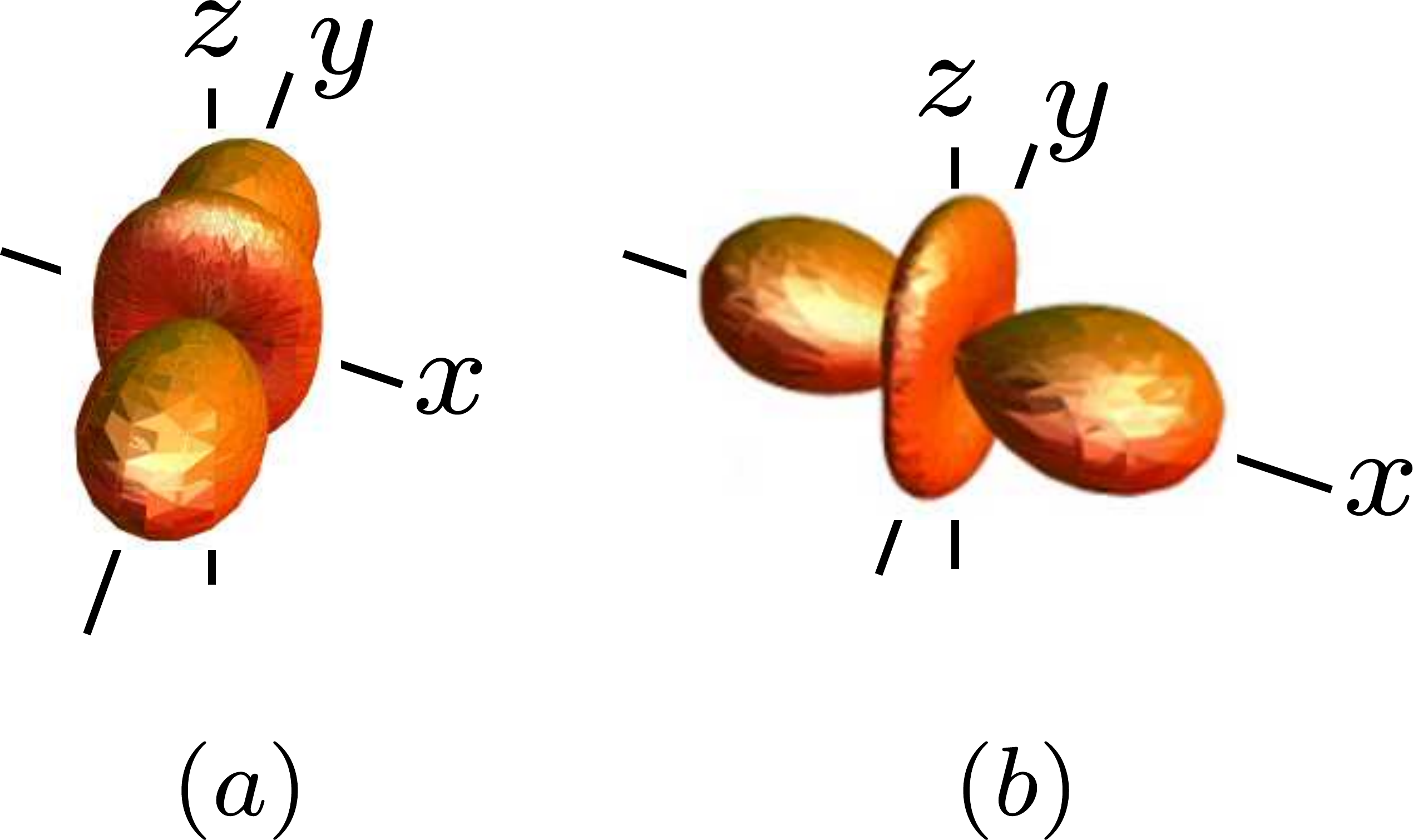}
 \end{center}
 \caption{(Color online.) The mixed orbital state for (a) orbital angle  $\theta=+\pi/2$ and  (b) orbital angle  $\theta=-\pi/2$.}
 \label{fig:orbitals}
 \end{figure}

 
 Energy minimization then yields $\theta=\pm \pi/2$, with the corresponding orbitals being the so-called mixed orbital ordering\cite{oles2000quantum,reitsma2005orbital, lee2012two,oles2000magnetic,vernay2004orbital}~: $|\theta=\pm\frac{\pi}{2}\rangle = \frac{1}{\sqrt{2}}|3z^2-r^2\rangle \pm \frac{1}{\sqrt{2}}|x^2-y^2\rangle$~.  This orbital ordering has a two-fold degeneracy shown in figure (\ref{fig:orbitals})), reflecting the $a=b$ axes symmetry of the tetragonally distorted spinel. 
 %=========================================================================================
 \subsection{Computation of spin exchange couplings}
The spin-orbital Hamiltonian (\ref{equn:hamiltonian}) shows that the dynamics of the orbitals and spins degrees of freedom are coupled to one other. This implies orbital ordering must dictate low temperature spin ordering.  Thus, once orbital-ordering is obtained, one can use the orbital-ordering angle $\theta=\pm\pi/2$ in the spin-orbital Hamiltonian in deriving an effective spin exchange interaction between spins in different crystallographic directions. In the three crystallographic planes ($XY$, $XZ$ and  $YZ$), the effective spin Hamiltonian has the form (upon neglecting the small $J_\sigma$ term in equn. (\ref{equn:hamiltonian})),
%\begin{align}
%H_{eff}^{spin} = (-J_\mu \langle\mathcal{W}_{ij}^{\alpha\beta}\rangle +J_\nu \langle\mathcal{V}_{ij}^{\alpha\beta}\rangle)\vec{S_i}.\vec{S_j}
%\end{align}
 \begin{eqnarray}
 H^{spin}_{xy}(\theta=\pm\pi/2)&=& 0.75 J_\mu ~\vec{S_i}.\vec{S_j}=J_{BB}~\vec{S_i}.\vec{S_j},\nonumber \\
  H^{spin}_{xz}(\theta=\pm\pi/2)&=&\pm 0.865\bar{\alpha}| J_\nu|~  \vec{S_i}.\vec{S_j}=\pm J'_{BB}~\vec{S_i}.\vec{S_j} \nonumber \\
   H^{spin}_{yz}(\theta=\pm\pi/2)&=& \mp 0.865\bar{\alpha}| J_\nu| ~ \vec{S_i}.\vec{S_j}=\mp J'_{BB}~\vec{S_i}.\vec{S_j}~. \nonumber \\
  \label{eqn:spininteraction}
 \end{eqnarray}
In obtaining the equations (\ref{eqn:spininteraction}), we have used the expression for the expectation values of the orbital operators $\mathcal{W}_{ij}$ (equation (\ref{eqn:expectation})) and $\langle \mathcal{V}_{ij}^{\alpha\beta}\rangle= \cos((\theta_i+\theta_j)/2+\chi_\gamma)\cos((\theta_i-\theta_j)/2)$, where $\chi_\gamma=4\pi/3,-4\pi/3$ and $0$ for $\gamma=x, y, z$ respectively and $\{\alpha, \beta, \gamma\}$ is a cyclic permutation of $\{ x, y ,z\}$.  For a system with tetragonal symmetry, the ratio of superexchange-based spin exchange couplings $\bar{\alpha} (\equiv J_\nu^{xz}(J_\nu^{yz})/J_\nu^{xy})<1$ . Our results show that the antiferromagnetic spin-exchange between the B-type moments (i.e., Mn$^{3+}$ ions) in the XY planes is given by $J_{BB}=0.75J_\mu$. However, the exchange coupling between B-type moments in the other planes is considerably weaker in value, $J'_{BB}=0.865\bar{\alpha}~|J_\nu| \ll J_{BB}$~, and can be either ferromagnetic or antiferromagnetic depending on the sign of the orbital angle $\theta$. 
%This calculations also reflect the fact that in the tetragonal phase $J_{BB}$  spin superexchange is much stronger than $J'_{BB}$. In a Mn$^{3+}$ tetrahedron, $XY$-planes spin interactions are antiferromagnetic for both orbital ordering whereas for a particular orbital ordering (i.e. either for $\theta=+\pi/2$ or $\theta=-\pi/2$) one of the vertical plane (i.e. either $XZ$- or $YZ$-plane) spin coupling is ferromagnetic and other is antiferromagnetic in nature. 
As a result, the geometrical frustration inherent in a Mn$^{3+}$ tetrahedron  is relieved by the mixed nature of the orbital-ordered ground state.
% This shows that one of the  non-planer interaction is ferromagnetic and other is antiferromagnetic in nature. This implies that after orbital get ordered no more frustration in the Mn$^{3+}$ spin tetrahedra, although it left behind two fold degeneracy.\par 
%  If we represent super-exchange constants in temperature  and consider J$_\tau=$1400K, J$_\mu=$7K then J$_{xy}=$5.25K and  J$_{xz/yz}=$0.374K , then according to mean field Ne\'el temperature would be $T_N=5.25\times 4 \times 2= 42K$ (for chain). Low values of the parameter due to lack of real value for Mn$_3$O$_4$ spinel. There could be  another contribution to the xy exchange from t$_{2g}$ superexchange. From Finer et.al. it has 24K contribution to the xy exchange. The parameter values for Mn$_3$O$_4$ are $J_{AA}$=4.9 K, $J_{BB}$=19.9 K, and $J_{AB}$=6.8 K. The strongest exchange coupling in Mn$_3$O$_4$ is amongst the Mn$^{3+}$ ions at the B sites\cite{srinivasan1983magnetic}. With these parameter values, we compute the T$_N= 160$K.
\subsection{Spin ordering in the Mn$_{3}$O$_{4}$ spinel}
 % At around 1400K, the Mn$_{3}$O$_{4}$ spinel system undergoes a structural phase transition from cubic to tetragonal (Jahn-Teller effect) and consequently there is orbital ordering ($\theta=\pm\pi/2$). Orbital-orbital interaction is very stronger than spin-orbital or spin-spin interaction, so it is clear that spin ordering will take place at very low temperature than orbital one.\par  
While the Mn$_{3}$O$_{4}$ spinel system undergoes a structural phase transition from cubic to tetragonal (and consequent orbital ordering) at 1400K, spin ordering happens only at around 42K. 
%So it is not clear why system remains in disorder phase through out that very large temperature range (1400K-42K). 
The reason for a lack of spin ordering in a temperature range as large as $42K < T < 1400K$ 
%There could be two reasons first, 
is indicated by the low values of the various spin exchange constants with respect to the inter-orbital coupling. 
%and corresponding Curie temperature is that very low (42K) and secondly, J$_{BB}$ is the strongest interaction among other and it is strong enough to sustain at high temperature (200K). 
The largest of these spin exchange coupling, $J_{BB}>0$, leads to the formation of one-dimensional antiferromagnetic spin chains in the [110] and $[1\bar{1}0]$ directions in this temperature range. While the Mn$^{3+}$ spins are $S=2$, the large on-site Hunds coupling ensures that only the $S^{z}=\pm 2$ is manifest in the dynamics. In effect, we can treat the spins as effectively spin-1/2, 
%As the chain interaction is AFM it could have 
with any 1D chain having two degenerate classical Ne\'{e}l-type ground-state configurations. 
%as shown in figure (\ref{fig:YKdegeneracy}) below.
\begin{figure}[h!]
\centering
      \includegraphics[scale=.2]{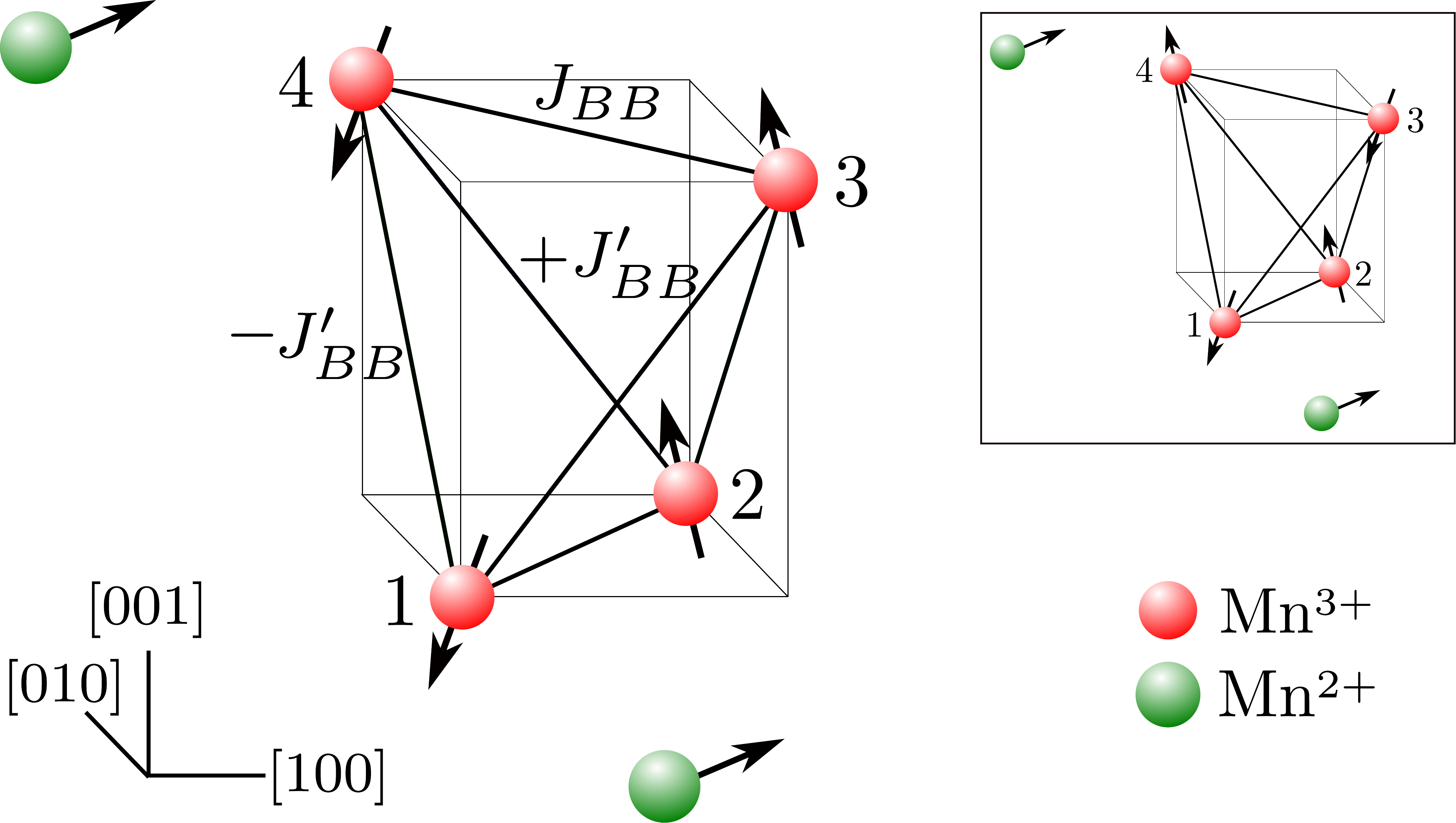}
  \caption{ (Color online.) Schematic diagram of a Yafet-Kittel ferrimagnetic spin configuration in Mn$_3$O$_4$ spinel. Two Mn$^{2+}$ spins align ferromagnetically along [110] direction whereas four Mn$^{3+}$ spin cant toward $[\bar{1}\bar{1}0]$ direction with a resultant moment along $[\bar{1}\bar{1}0]$. The inset shows another Yafet-Kittel configuration which is degenerate with that shown in the main figure.}
  \label{fig:YKdegeneracy}
\end{figure}
As is well known, such 1D spin chains cannot display any true long-ranged ordering of spins at any finite temperature; at best, a quasi-long ranged (or algebraic) order is obtained \cite{fradkin2013field}. Thus, a true long-ranged spin ordering can only be obtained at low-temperatures through weaker spin exchange couplings, including the antiferromagnetic Mn$^{3+}$-Mn$^{2+}$ coupling ($J_{AB}$) and the inter-chain spin exchange couplings $J'_{BB}$ (some of which are ferromagnetic, and some antiferromagnetic, in nature)~\cite{srinivasan1983magnetic,chartier1999ab}.
%t high temperature this two configurations could tunnel (thermally or quantum mechanically) and resultant there will be no spin ordering. Now as temperature goes down J$'_{BB}$ exchange value become compatible and forms a there dimensional spin complex. 
\par
In order to understand the appearance of the ferrimagnetic Yafet-Kittel (Y-K) long-ranged spin order at low-temperatures, we will consider the competing interactions within a single Mn-ion complex of 2 Mn$^{2+}$ spins and a tetrahedron of 4 Mn$^{3+}$ spins. The canting of the Mn$^{3+}$ spins in the Y-K state can be seen to arise from having to satisfy the weak antiferromagnetic Mn$^{3+}$-Mn$^{2+}$ coupling ($J_{AB}$) together with the in-chain Ne\'{e}l order due to $J_{BB}$. The yet weaker spin exchange couplings $J_{BB}^{'}$ and $J_{AA}$ help relieve the frustration and stabilise the Y-K ground state \cite{yafet1952antiferromagnetic, menyuk1962classical, lyons1960method,willard1999magnetic}. Note that the antiferromagnetic nature of $J_{AB}$, together with the canting of the Mn$^{3+}$ spins, will lead to a ferrimagnetic ground state: the 4 Mn$^{3+}$ spins will possess a total magnetic moment smaller than, and anti-aligned with, the 2 Mn$^{2+}$ spins.  
%In a Mn$^{3+}$ tetrahedron, one can subdivide the four spins into two groups according to the nature of their interactions. For instance, spins (1,4) and spins (2,3) form two groups for the tetrahedron shown on the left-hand side in figure (\ref{fig:YKdegeneracy}). Note that the intra-group interaction is ferromagnetic (FM) in nature, while the inter-group interaction is antiferromagnetic (AFM) in nature. J$_{BB}$ is strongest interaction than other superexchange constant J$_{AB}$(interaction between Mn$^{3+}$ and Mn$^{2+}$spin ), J$_{AA}$ (interaction between Mn$^{2+}$-Mn$^{2+}$  spin) and  J$'_{BB}$, which clearly indicate that low temperature spin model for these spinel falls into Yafet-Kittel(YK) spin model \cite{yafet1952antiferromagnetic, menyuk1962classical, lyons1960method} which support triangular ferrimagnetic spin ordering. Since $J_{AB}$ and $J_{BB}$ both antiferromagnetic interaction then for strong enough $J_{BB}$   spins in Mn$^{3+}$ tetrahedron rotate away from purely antiferromagnetic configuration with the Mn$^{2+}$ spins\cite{willard1999magnetic}.
% Due to a$=$b symmetry YK spin configurations have two degenerate partner as shown in figure.
%\par 
As the system has tetragonal ($a=b$) symmetry as well as orbital degeneracy, the Y-K spin ordering is doubly-degenerate (see figure (\ref{fig:YKdegeneracy})). 
%and with both configuration possessing the same resulting component of spin and antiparallel to A-spin).
 %Due to this spin-orbital-lattice instability system will under goes complicated spin ordering as temperature goes down(like incommensuration, spin spiral, commensurate cell doubling etc.). %which may be explain by Dzyloshinskii-Moriya type interaction). 
It is also known that an orthorhombic structural distortion finally stabilises magnetic and orbital ordering in the system at around $32K$. 
%Due to this distortion spins and orbitals will choose a particular state. 
%and system settle to one of YK-spin ordering. As system has orbital as well as spin degeneracy there could be two types of instability namely orbital-lattice and spin-lattice. As the system has $\theta=\pm \pi/2$ orbital degeneracy it could couple with lattice, resultant another Jahn-Teller types structural distortion. Where is for 
Thus, in the next section, we show how a spin-lattice coupling lifts the spin-degeneracy via a orthorhombic distortion of the 
%lattice structure  the 
Mn$^{3+}$ tetrahedra~\cite{chung2013low, nii2013interplay, tchernyshyov2011spin}, attaining a cell-doubled Y-K spin-ordered ground state. 
%So from tetragonal to orthorhombic distortion could be out come of either one of the two effect for Mn$_3$O$_4$ spinel.
 \section{Spin-lattice interaction in the spinel} 
Although the geometrical frustration of the spinel lattice is relieved by the tetragonal distortion, a two-fold degeneracy remains in the choice of the nature of the exchange interactions between spins in the $XZ$ and $YZ$ planes, J$'_{BB}$, in a given tetrahedron. This can be observed in equation (\ref{eqn:spininteraction}) above. As shown in Fig.(\ref{fig:orthoforce}) below, these two degenerate configurations are associated with different orthorhombic (a$\neq$b) lattice distortions of the system~\cite{chung2013low}. Studies of the effects of a spin-lattice coupling in spinel~\cite{PhysRevB.89.134402, matsuda2007spin} and pyrochlore~\cite{tchernyshyov2011spin} systems show how magnetic ordering can be achieved via orthorhombic distortions. In seeking the eventual stabilisation of magnetic order in orthorhomically-distorted Mn$_{3}$O$_{4}$, we follow an analysis similar to Ref.(\cite{tchernyshyov2011spin}).
\par 
 Allowing for a dependence of the superexchange constant $J_{ij}$ on the distance between spins at sites $i$ and $j$, the contribution of such a pair to the exchange energy \cite{ PhysRevLett.85.4960,tchernyshyov2002spin, PhysRevLett.88.067203} is given by
\begin{equation}
 E_{ij} =[J+(dJ/dr)\delta r_{ij}+.....](\vec{S}_i\cdot \vec{S}_j)~.
\end{equation}
 Therefore, the spins exert on each other a force, $-(dJ/dr)(\vec{S}_i\cdot \vec{S}_j)$, which is attractive or repulsive depending on the angle between the two spins. In general, the angles between the nearest-neighbour spins placed at the vertices of a lattice can be unequal. This leads naturally to an unbalanced force acting on each spin, cooperatively resulting in a deformation of the lattice. Such a spin-lattice coupling has been called the ``spin-Teller" interaction~\cite{matsuda2007spin,tchernyshyov2002spin}, and whose form is taken be~\cite{lacroix2011introduction}
 \begin{eqnarray}
 \mathcal{H}_{SL}=\sum_{i,j} (\partial J_{ij}/\partial r)(\vec{S}_i\cdot \vec{S}_j)\delta r_{ij}~,
 \end{eqnarray}
where $\delta r_{ij}$ is the elongation along the one axis (a or b), and $\partial J_{ij}/\partial r$ is itself a negative quantity. 
\par
\begin{figure}[h!]
 \centering
 \includegraphics[scale=.25]{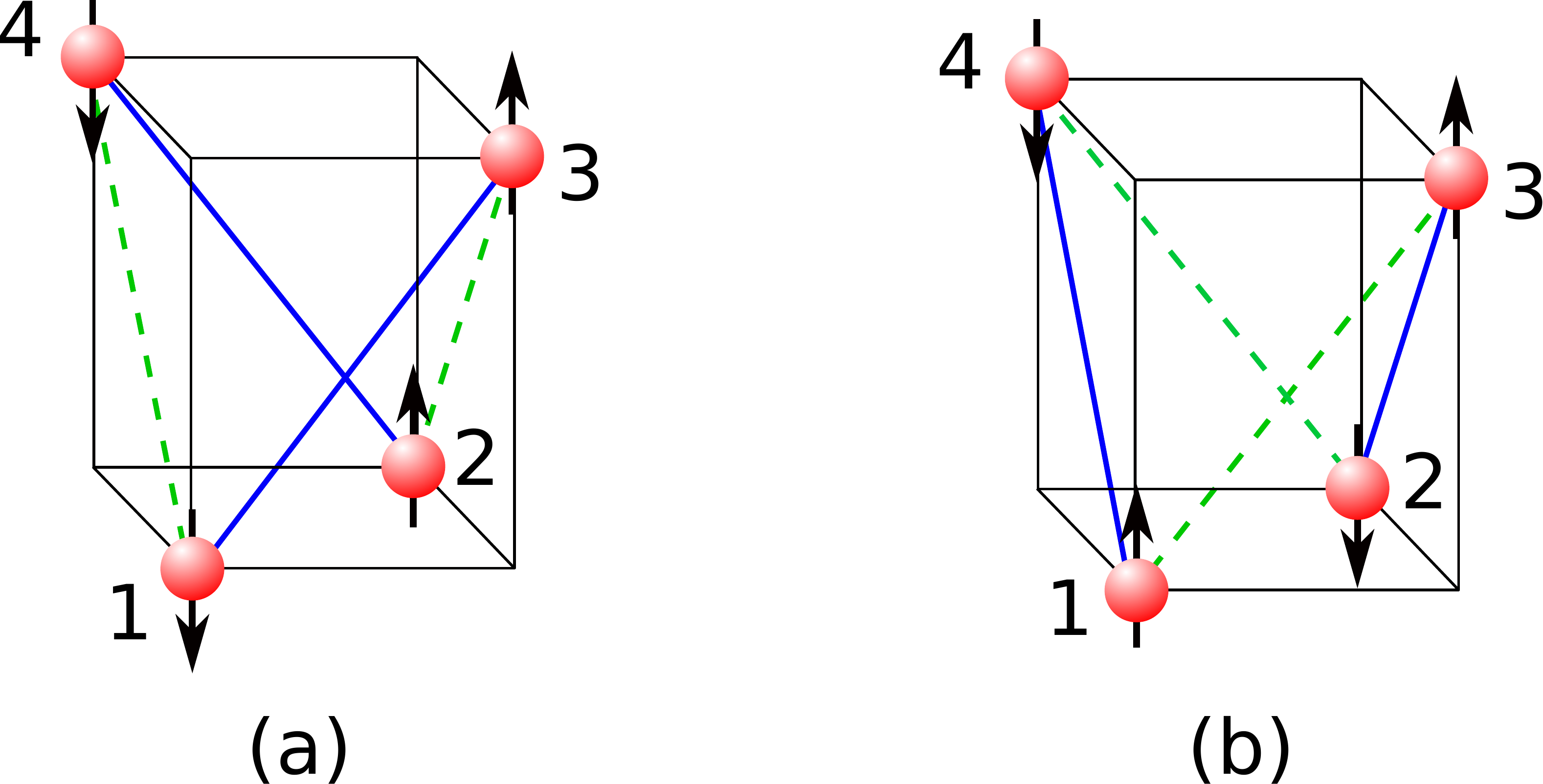}
 \caption{ (Color online.) The two orthorhombic forces given in equation (\ref{orthfor}). Dotted lines correspond to the repulsive forces between spins and the resultant elongation of the plane on which they are lying, while solid lines correspond to attractive forces and resultant compression of the plane. (a) This configuration support elongation along y-axis while (b) supports elongation along x-axis.}
 \label{fig:orthoforce}
 \end{figure}
 Due to inter-orbital interactions, tetrahedra of Mn$^{3+}$ spins already tetragonally distorted (i.e., have a reduced symmetry, $c > a = b$). From the spin-orbital Hamiltonian, we know that the spin-exchange couplings in the $XY$ planes is strong and antiferromagnetic. This makes it plausible that, among all possible phonon modes for a tetrahedron, a weak spin-lattice coupling will likely lead to an orthorhombic distortion. The form of the orthorhombic force is given by~\cite{tchernyshyov2002spin}
\begin{eqnarray}
f &&= (\vec{S}_1-\vec{S}_2)\cdot(\vec{S}_3-\vec{S}_4)/2 \nonumber \\ &&\text{or} ~~  (\vec{S}_1-\vec{S}_2)\cdot(\vec{S}_4-\vec{S}_3)/2~,
 \label{orthfor}
\end{eqnarray}
where the distortion corresponding to the first term is shown in figure (\ref{fig:orthoforce}(a)) and that corresponding to the second in figure (\ref{fig:orthoforce}(b)). The energy of the system then has the form
\begin{equation}
E= -J'f\cdot\delta r + k \delta r^2/2 ~,
\end{equation}
 where $\delta r$ is the amplitude of orthorhombic distortion,  J$'$ ($=\partial J/\partial r$) and k are the magneto-elastic and elastic constants respectively.
 \par
\begin{figure}[h!]
 \centering
 \includegraphics[scale=.055]{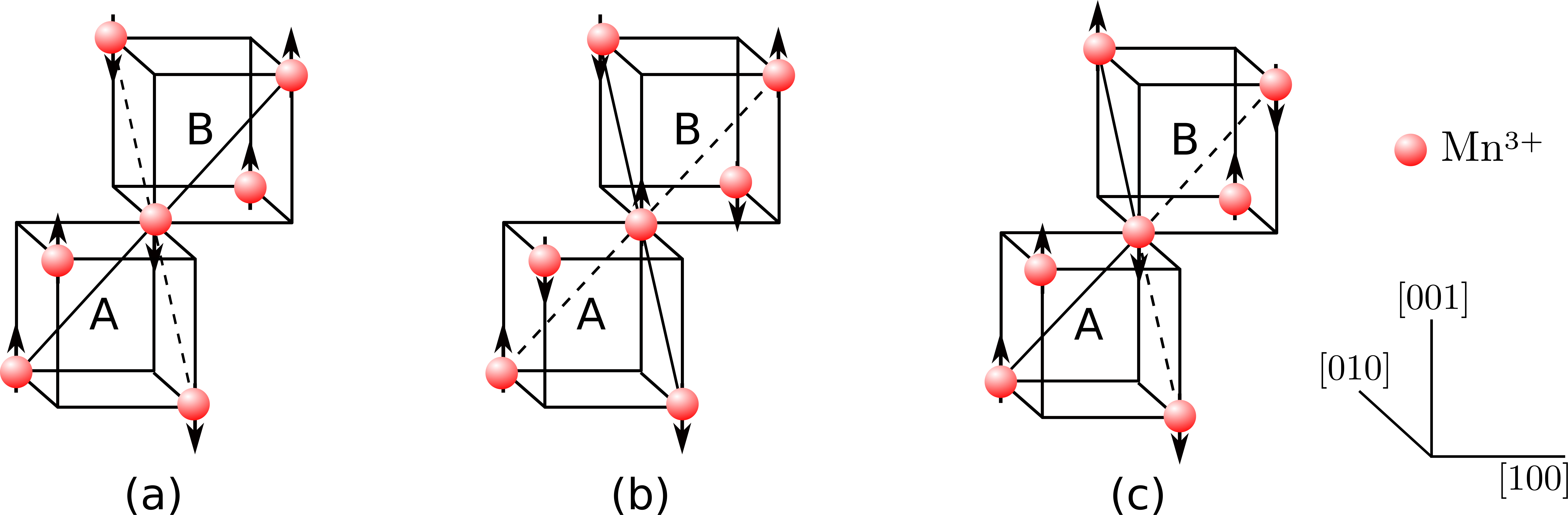}
 \caption{ (Color online.) Two nearest-neighbour tetrahedra A and B in a spinel lattice with their respective orientations of the Mn$^{3+}$ ions. In all three configurations, the XY plane interaction is strong and antiferromagnetic. The spin-lattice coupling leads to an overall distortion along [010] direction in figure (a), along [100] direction in figure (b), and along the $[110]$ direction in figure (c) leading to a cell-doubled Yafet-Kittel spin state.}
 \label{fig:spin-lattice}
 \end{figure}

 
 Now to consider the collective orthorhombic distortion ($q=0$), we have to take into account the fact that the relevant degree of freedom arises from the existence of two different types of tetrahedra in a spinel, namely A and B (as shown in figure (\ref{fig:spin-lattice})), that differ in their orientation.
 At linear order in the displacements, there are only two active modes which couple to the spins -- $E_g$(acoustic mode, overall orthorhombic distortion) and $E_u$ (optical mode, tetrahedrons distorted in exactly opposite direction). These two modes can be expressed in terms of the scalar quantities
 \begin{eqnarray}
 Q^g=\frac{Q^A+Q^B}{\sqrt{2}},\hspace*{.5cm} Q^u=\frac{Q^A-Q^B}{\sqrt{2}}~.
 \end{eqnarray}
The spin -lattice energy is then
 \begin{eqnarray}
 E(f^A,f^B,Q^A,Q^B)=J'(Q^A.f^A&&+Q^B.f^B)+\frac{k_g}{2}|Q^g|^2\nonumber \\ &&+\frac{k_u}{2}|Q^u|^2~.
 \end{eqnarray}
The minimisation of this energy with respect to the two lattice modes $Q^g$ and $Q^u$ we have 
 \begin{eqnarray}
 E(f^A,f^B)=&& -\frac{(K_g+K_u)(|f^A|^2+|f^B|^2)}{4}\nonumber \\ &&-\frac{K_g-K_u}{2} f^A.f^B~,
 \label{eqn:spin-lattice}
 \end{eqnarray}
 where $K_{g/u}=J'^2/k_{g/u}$ are the effective spin-lattice coupling constants. The first term in the expression of the energy is the self-interaction term for individual tetrahedra, and second term is the interaction term between tetrahedra.
 \par 
 In the limit $K_g > K_u$ (i.e., E$_g$ mode softens first), the distortion of the two tetrahedra have the same nature, i.e., in order to minimise energy, $f^A$ and $f^B$ have to support a distortion along same axis (see  figure(\ref{fig:spin-lattice}(a), \ref{fig:spin-lattice}(b)). However, in the limit $K_u > K_g$ (i.e., E$_u$ mode softens first), the two tetrahedra distort in an equal and opposite manner, i.e., $f^A$ and $f^B$ elongate along x- and y-axis respectively (see figure (\ref{fig:orthoforce})). Clearly, this case leads to a doubling of the magnetic unit-cell, as observed in various experiments on Mn$_{3}$O$_{4}$~\cite{suzuki2008magnetodielectric,kim2010mapping,kim2011pressure,nii2013interplay,chung2013low}.
There remains one subtlety worth discussing. We recall that our analysis yielded two degenerate configurations of the orbitals (characterised by the mixing angle $\theta=\pm\pi/2$) in the tetrahedral phase. This immediately leads to two degenerate possible cell-doubled structures: each of the two-sublattices can have an orbital state with either $\theta=\pi/2$ or $\theta=-\pi/2$, with an alternation leading to the doubling of the unit cell. A spontaneous symmetry breaking mechanism will, of course, be needed to choose between these two cell-doubled structures. This mechanism can, for instance, arise from the choice of the magnetic easy-axis in the system. In Mn$_{3}$O$_{4}$, this is due to the Mn$^{2+}$ spins. Here, if the easy axis is along the $[110]$ direction, the orbital angle of cell A is $\theta=\pi/2$ and cell B is $\theta=-\pi/2$ (see figure(\ref{fig:spin-lattice})(c)), while for the easy axis being $[1\bar{1}0]$ the orbital angles of the two sublattices are exchanged.
 %==================================================================================
\subsection{Comparison with experiments performed at low-temperature}
%\emph{Spin-lattice perspective}: 
We will now attempt a comparison of the results obtained from our theoretical analysis thus far with the experimental observations on Mn$_{3}$O$_{4}$ at low temperatures. The experiments probe the interplay of a magnetoelastic coupling and external magnetic fields on the magnetic and structural ordering of the system~\cite{kim2010mapping,kim2011pressure,nii2013interplay}. We begin by recalling that for $T<32K$, the ordered magnetic state involves a cell-doubled state of distorted tetrahedra (as discussed above). When an external magnetic field is applied along the [100] direction with $T<32K$, Nii et.al.~\cite{nii2013interplay} observe a transition to a state with uniformly distorted tetrahedra. This can be understood as follows. The application of the field, $-B_x(S_i^x+S_j^x)$, will naturally mean that spins in the XZ plane will favour alignment along the [100] direction. This ferromagnetic alignment in the XZ planes spins ($\vec{S}_i.\vec{S}_j=+1$) supports a uniform odering of the orbitals with the orbital angle $\theta=-\pi/2$ in equation (\ref{eqn:spininteraction}) (see figure(\ref{fig:orbitals})). Further, such uniformly distorted tetrahedra will cause a softening of the acoustic $E_{g}$ mode (see equation (\ref{eqn:spin-lattice})), eventually overturning the cell-doubled ground state. Similarly, when the B-field is applied along the [010] direction, spin alignment is favoured along the y-axis ($-B_y(S_i^y+S_j^y)$). In turn, the ferromagnetic alignment in the YZ planes ($\vec{S}_i.\vec{S}_j=+1$) supports a uniform orbital ordering with  $\theta=\pi/2$ (see figure(\ref{fig:orbitals})). Once again, the magnetoelastic coupling ensures that the cell-doubled state will eventually be replaced with a state comprised of uniformly distorted tetrahedra. This is again in keeping with the observations of Ref.(\cite{nii2013interplay}).
%\par 
%If B-field is along [110] direction, spins will want to polarized equally along x and y direction. This means that either case above are equally favoured. But the orbitals are director spins and can not be rotate in xy plane by the B-field. Choice has to be made for orbital at each site with global constraint  that equal shearing by the two orbitals takes place. Then the simplest possibility is a cell-doubled structure. B-field in [110] will want a cell-doubled structure which is stretched along [110]. Which choose $\theta=-\pi/2$ orbital at site-I and $\theta=\pi/2$ on site-II. On the other hand, when B-field in [1$\bar{1}$0] will want a cell-doubled structure which is stretched along [1$\bar{1}$0]. In this case site-I preferred $\theta=\pi/2$ and site-II $\theta=-\pi/2$ orbital.\par
%This cell-doubling  can happen via a softening of the optical mode (equation (\ref{eqn:spin-lattice})): 
%\begin{align}
%k_u\rightarrow 0 \Rightarrow K_u\rightarrow \infty \Rightarrow f^A\cdot f^B=-1
%\end{align}
%Finally if we start from [110]-favour state and apply [1$\bar{1}$0] aligned B-field, we will go from configuration-I to configuration-II, through an intermediate undistorted state (tetragonal). Tunnelling could do that job at low temperature.
\begin{figure}[h!]
 \centering
 \includegraphics[scale=.5]{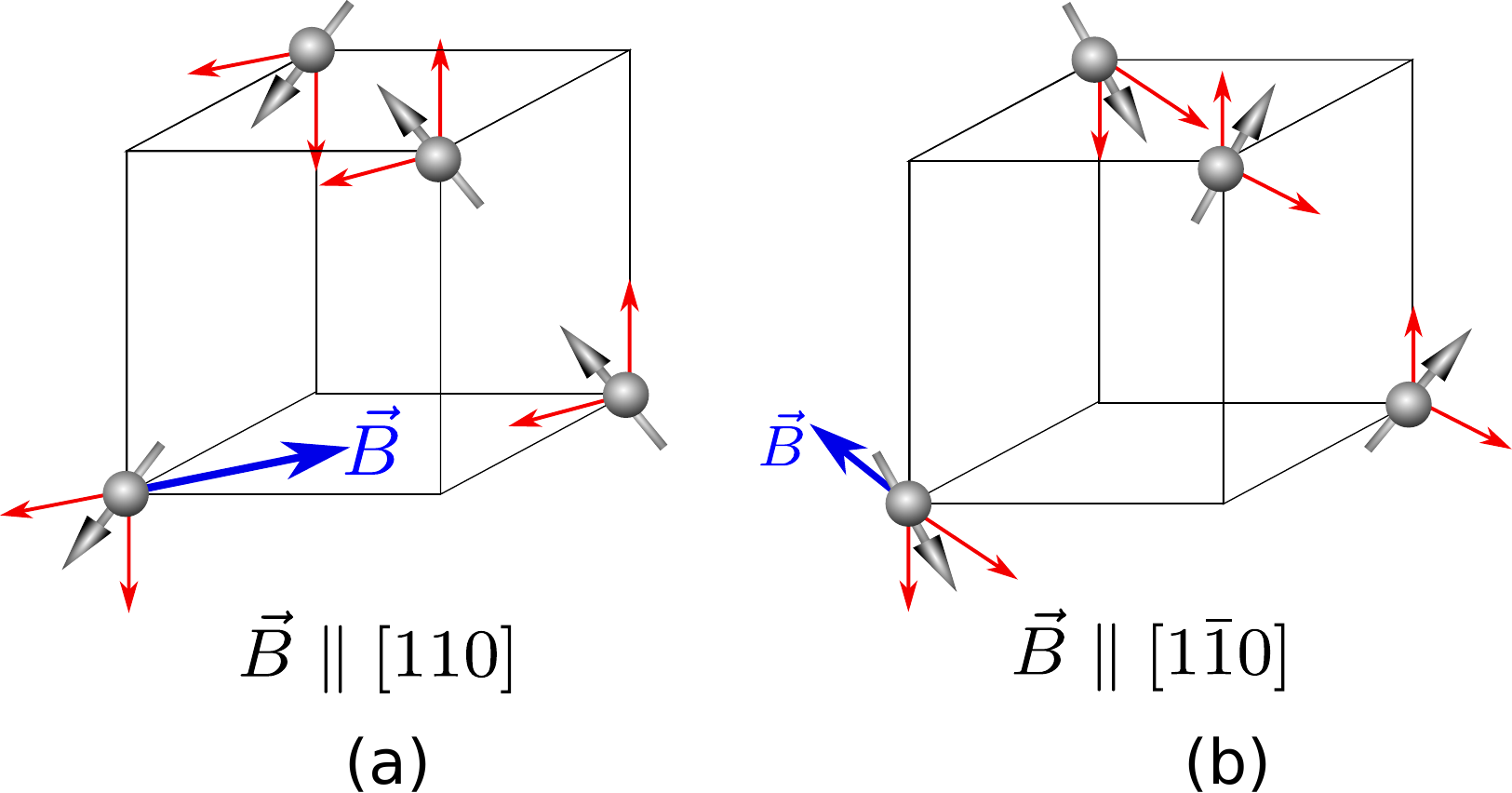}
 \caption{(Color online.) (a) Application of a magnetic field ($\vec{B}$) along the [$1 1 0$] easy-axis enhances the spin-lattice interaction even for $T>32K$, and leads to an elongation along the same direction. Red arrows show the spin components along the $c$ and easy axes. (b) Application of a sufficiently large magnetic field along the [$1 \bar{1} 0$] direction in the orthorhombically distorted phase below $32K$ leads to a rotation of the easy axis towards [$1 \bar{1}0$]. The spin-lattice interaction will then lead to an elongation along this direction. Red arrows show the spin components along the $c$ and [$1 \bar{1}0$] directions.}
 \label{fig:spinlatticefield}
 \end{figure}
 \par 
The experiments by Kim \emph{et. al.}~\cite{kim2010mapping,kim2011pressure} show, on the other hand, that for temperatures $32K < T<42K$, the application of a magnetic field greater than 1 Tesla along the magnetic easy axis (the [110] direction, due to the Mn$^{2+}$ spins) leads to the lattice being distorted from tetragonal to orthorhombic. Insight into this finding can again be gained as arising from an increase of the spin-lattice interaction resulting from the applied external magnetic field. The magnetization will now be maximum when the magnetic moments align along the $[\bar{1}\bar{1}0]$ direction. 
%If there is no resultant magnetic moments due to Mn$^{3+}$ spins along [$\bar{1}\bar{1}0$] direction, no reduction of Mn$^{2+}$ moments along [110] direction. 
While the Yafet-Kittel configuration involves a canting of the Mn$^{3+}$ spins away from the easy axis ([110]), the application of a sufficiently large external magnetic field along the easy axis will cause them to align along the $[\bar{1}\bar{1}0]$ direction (see figure \ref{fig:spinlatticefield}(a)). 
%Mn$^{3+}$ spins to align co-linearly along [001] direction This gives a maximisation of the spin overlap $\vec{S}_i.\vec{S}_j$ in the $[\bar{1}00]$ and $[0\bar{1}0]$ directions. 
In turn, this leads to an increased spin-lattice interaction in equation (\ref{eqn:spin-lattice}), resulting in an orthorhombic distortion even for $T>32K$. 
%So we conclude here, external magnetic field along [110] increases spin-lattice coupling and lattice distorted from tetragonal to orthorhombic. 
%the form of the Hamiltonian in presence of external magnetic field will be
% \begin{align}
% H=J'\langle I^\alpha I^\beta\rangle(\vec{S}_i\cdot \vec{S}_j)\delta_{ij}-\vec{h}\cdot(\vec{S}_i+ \vec{S}_j), \qquad \text{h is the strangth of external magnetic field.}
% \end{align}
% \textcolor{red}{There could be one alternative explanation about parallel field distortion as follow. Above 33K as a$ = $b orbital degeneracy $ (\theta=\pm \pi/2) $ remain intact. External magnetic field mixed $d_{xy}$ orbital with degenerate orbital via spin-orbit interaction induce by magnetic field, and due to  same phase of these orbitals electron density increase along [110] direction and resultant distortion along these direction. These explanation can also explain distortion along [1$\bar{1}$0] when transverse magnetic field apply along easy axis. See figure(\ref{fig:mixing})}.
\par 
% \begin{figure}[h!]
% \centering
% \includegraphics[scale=.2]{orbitalmixing}
% \caption{The nature of the mixed orbital configuration is shown here.}
% \label{fig:mixing}
% \end{figure}
Further, Kim \emph{et al.} also observe that by applying the magnetic field along the $[1\bar{1}0]$ direction (i.e, a field transverse to the [110] easy axis of the system) for $T<32K$, a structural transition takes place towards an orthorhombic distortion along the $[1\bar{1}0]$ direction as the applied field is gradually increased. At intermediate values of the transverse field, the system appears to be in a tetragonal phase with a loss of the magnetic ordering. These findings can be understood by noting that the transverse field rotates the choice of the easy axis from its natural direction ([110]) to its perpendicular direction ($[1\bar{1}0]$) (see figure \ref{fig:spinlatticefield}(b)). The passage from one to the other will also involve a change from one cell-doubled state of orthorhombically distorted tetrahedra (cell A with orbital angle $\theta=\pi/2$, cell B with $\theta=-\pi/2$) to another in which the orbital angles for the two sublattices are exchanged. As discussed above, both configurations are resultant from the model for the magnetoelastic coupling. Further, at the transition, fluctuations between the two orthorhombic configurations are large and we can expect the outcome to be a tetragonal phase for the system in which the Yafet-Kittel spin order is lost. 
%The reason behind this is that, in the leading order of spin-lattice interaction it support distortion along both x or y axis and higher order term lift the degeneracy and chooses one direction let's say x-axis. Resultant a$>$b tetragonal to orthorhombic distortion took place. Now transverse field try to mixed up these two distortion and for certain value of field system return to the tetragonal phase and no magnetisation. For high enough transverse field magnetic field bits higher order spin-lattice term  and choose other direction (y-direction) for orthorhombic distortion along with magnetization along $[1\bar{1}0]$ direction. 
%The Hamiltonian could be written as 
% \begin{align}
% H=J'\langle I^\alpha I^\beta\rangle(\vec{S}_i\cdot \vec{S}_j)\delta_{ij}-\vec{h}_\bot\cdot(\vec{S}_i+ \vec{S}_j)
% \end{align}
% \par 
% When field is applied along [100] direction then $\theta=-\pi/2$ orbital will be occupied which support ferromagnetic ordering in xz plane and antiferromagnetic along yz plane. This configuration support overall distortion along x axis (a$>$b distortion). Whereas when field applied along [010] direction  $\theta=\pi/2$ orbital will favour  and overall distortion along y axis. 
  
 \section{Discussion and summary}
We have, in this work, developed and analysed a microscopic model for the understanding of orbital and spin ordering in the Mn$_3$O$_4$ spinel. Our analysis demonstrates that the qualitative properties of the spinel can be explained by a spin-orbital Hamiltonian derived by a careful consideration of the 90$^o$ nature of the Mn-O-Mn bonding angle, as well as the different electronic energy states of the Mn$^{3+}$ ion. In this way, we realised that the charge transfer (or Goodenough) superexchange process is of greater importance than the direct (Anderson) superexchange for the 90$^o$ metal-ligand-metal bond (see, for instance, equn.(\ref{eqn:Goodenoughprocess})). Upon considering superexchange interactions together with Jahn-Teller orbital-lattice coupling, we find that inter-orbital interactions are much stronger than spin-orbital as well as spin-spin interactions. This indicates that orbitals will likely order at energy scales much higher than the spins. Further, the nature of the orbital ordering will influence the nature of the spin ordering. %Another important aspect is that the intra-orbital part of the spin-orbital Hamiltonian derived from purely electronic SE processes has the  same  form as the Jahn-Teller Hamiltonian derived from phonon-orbital interaction for this spinel. 
\par
A consequence of the 90$^{o}$ bonding angle is that the interaction energy between a pair of Mn$^{3+}$ ions is less for similar orbitals ($\theta_i\simeq\theta_j$) than dissimilar orbitals ($\theta_i\simeq\theta_j+\pi$). With this in mind, a variational analysis of our Hamiltonian yields a mixed orbital configuration in the tetragonal phase of the spinel. This mixed ordering of the orbitals then plays an important role in relieving the geometric frustration of spins in a Mn$^{3+}$ tetrahedron. In the tetragonal phase, small values of the effective spin exchange couplings lead to spins ordering along one-dimensional chains along the [110] and  $[\bar{1}10]$ directions, and the experimentally observed three-dimensional Yafet-Kittel (Y-K) type spin ordering is delayed till very low temperatures. The canted nature of the spins in the Y-K state of Mn$^{3+}$ tetrahedra arise from the competing weak antiferromagnetic Mn$^{3+}$-Mn$^{3+} ~(J_{BB})$ and Mn$^{2+}$-Mn$^{3+} ~(J_{AB})$ spin exchange couplings. %have same nature  (which is AFM)  and almost same order of magnitude Mn$^{3+}$ spins canted toward $[\bar{1}\bar{1}0]$ direction, 
This results in the ferrimagnetic Y-K spin ordering of the Mn$_3$O$_4$ spinel system. 
%In tetragonal phase down to low temperature system posses degeneracy ($a=b$). 
Further, we have shown that a spin-lattice coupling finally lifts the two-fold degeneracy of the Y-K spin configuration via an orthorhombic distortion, leading to a cell-doubled Y-K state. 
%In collective mode optical mode of vibration distort whole lattice to get cell double Y-K spin ordering in orthorhombic phase. 
Our model for the spin-lattice interaction is also able to explain the experimentally observed effects of an external magnetic field effect on the Y-K state~\cite{kim2010mapping,kim2011pressure,nii2013interplay,chung2013low}.
\par
We end by outlining some open directions. Kim \emph{et al.} show that a cell-doubled orthorhombic phase with the magnetic easy-axis along the [110] direction can be manipulated using a magnetic field leading to a similar phase with the easy-axis along $[\bar{1}10]$, with an intermediate tetragonal phase in which the Y-K spin order is lost~\cite{kim2011pressure}. It remains an open question on whether such a fluctuation-induced tetragonal phase can be stabilised at yet lower temperatures using, for instance, pressure. As suggested by a recent experiment on the perovskite magnetic insulator KCuF$_{3}$~\cite{yuan2012}, a quantum orbital-spin liquid state could potentially be realised in Mn$_{3}$O$_{4}$. Further, experimental signatures of the mixed nature of the orbital state may be possible to detect in, for instance, X-ray measurements carried out in the low-temperature orthorhombic phase~\cite{lee2012two}. Finally, in a future work, we will present results on the orbital and spin excitation spectra in various ordered phases; this could also be useful in tracking the passage between competing ground states.
% In summary, we presented a derivation of spin-orbital Hamiltonian for Mn$_3$O$_4$ spinel system. To get orbital ordering we variationally analysis intra-orbital part of the Hamiltonian only. Mixed orbital ordering relived geometrical frustration in the Mn$^{3+}$ tetrahedron. Finally, spin-lattice interaction removed all degeneracy present in the system.
\section*{Acknowledgements}
The authors thank S. Jalal, A. Mukherjee, V. Adak, N. Bhadra, A. Basu, R. Ganesh, G. Dev Mukherjee, S. L. Cooper, R. K. Singh, C. Mitra, Y. Sudhakar and K. Penc for several enlightening discussions. S. Pal acknowledges CSIR, Govt. of India and IISER-Kolkata for financial support. S. L. thanks the DST, Govt. of India for funding through a Ramanujan Fellowship (2010-2015) during which a substantial part of this work was carried out.
  %==================================================================================
\section*{Appendix\label{appendix}}

In this appendix, we present the detailed contributions to various superexchange (SE) couplings from the ``M'', ``O'' and ``N'' configurations in addition to that presented in section \ref{sec:spinorbitalHam}
\begin{widetext}
 \begin{eqnarray} 
\left[M,\uparrow\downarrow;\uparrow\uparrow\right] &=&  2[TtD]+2[TtS] \nonumber\\
\left[M,\uparrow\uparrow;\uparrow\downarrow\right] &=& [TtD]+[TsD]+[TtS]+[TsS] \nonumber\\
\left[ M,\uparrow\uparrow;\downarrow\downarrow \right] &=& \frac{2}{3}\bigr([TtD]+[TtS]+[DtD]+[DtS]+[GtD]+[GtS]\bigl) \nonumber\\
\left[O,\uparrow\uparrow;\downarrow\downarrow\right] &=& [DtD]+2[DtS]+[StS]\nonumber\\
\left[O,\uparrow\downarrow;\downarrow\uparrow\right] &=& \frac{1}{2}\bigr([DtD]+[DsD]+2[StD]+2[SsD]+[StS]+[SsS]\bigl)\nonumber\\
\left[N,\uparrow\uparrow;\uparrow\uparrow\right] &=& 4[TtT]\nonumber\\
\left[N,\uparrow\uparrow;\uparrow\downarrow\right] &=& \frac{2}{3}\bigr([TtT]+[TsT]+[TtD]+[TsD]+[TtG]+[TsG]\bigl)\nonumber\\
\left[N,\uparrow\uparrow;\uparrow\downarrow\right] &=& \frac{2}{3}\bigr([TtT]+[TsT]+[TtD]+[TsD]+[TtG]+[TsG]\bigl)\nonumber\\
\left[N,\uparrow\uparrow;\downarrow\downarrow\right] &=& \frac{4}{9}\bigr([TtT]+[DtD]+[GtG]+2[TtG]+2[DtG]+2[TtD]\bigl)\nonumber\\
\left[N,\uparrow\downarrow;\downarrow\uparrow\right] &=& \frac{2}{9}\bigr([TtT]+[TsT]+[DtD]+[DsD]+[GtG]+[GsG]\nonumber \\ && +2[TtD]+2[TsD]+2[TtG]+2[TsG]+2[DtG]+2[DsG]\bigl)\nonumber\\
\left[N,\uparrow\downarrow;\downarrow\downarrow\right] &=& \frac{4}{3}\bigr([TtT]+[DtT]+[GtT]\bigl)
\end{eqnarray}
\end{widetext}
Using above equations, one can write the various superexchange constants ($K$) in equation (\ref{eqn:SEconstant}). For the triplet configuration at metal ions sites, this gives
\begin{equation}
K_L^T = \sum_{\sigma,\sigma'}[L,\uparrow\uparrow;\sigma \sigma']~,
\end{equation}
while for the singlet configuration, we obtain $K_L^S=2K_L^{\uparrow\downarrow}-K_L^T$~, where $\sigma,\sigma'\in \{\uparrow,\downarrow \}$. In obtaining a more compact form of the Hamiltonian (\ref{eqn:spinorbitalHamiltonian}), one has to put the expression for various projection operators in the expression. The final form of the spin-orbital Hamiltonian is shown in equation (\ref{equn:hamiltonian}) with superexchange constants given by
\begin{widetext}
 \begin{eqnarray}
J_\tau &=& \frac{1}{2}(J_O^O-2J_M^O+J_N^O)\nonumber \\
&=& \frac{1}{18} \Bigg[ \bigg\{ 9(SS)_+ + 4(GG)_+ + (DD)_+ +6(DS)_+ + 18\{TG\}_+ + \frac{456}{10}[TtT]+\frac{184}{10}[TsT]\bigg\} \nonumber \\
&& \qquad\qquad - \bigg\{12(GS)_+ + 9\{TD\}_+ + 27\{TD\}_+ + 4(GD)_+\bigg\} \Bigg]~,
\end{eqnarray}

%===================================================================================
%And
\begin{eqnarray}
J_\sigma &=& -\frac{1}{2}(J_O^S+2J_M^S+J_N^S)\nonumber \\
&=& -\frac{1}{18}\Bigg[\bigg\{21(DD)_- + 9(SS)_- + 16(TT)_- + 30(DS)_- + 20(DG)_- + 12(GS)_-\bigg\}\nonumber \\
&& \qquad\qquad - \bigg\{40(TD)_- + 24(TS)_- + 16(TG)_- + 4(DD) + 4(GG)\bigg\}\Bigg]~,
\end{eqnarray}
%==================================================================================
\begin{eqnarray}
J_\nu &=& -\frac{1}{2}(J_O^S-J_N^S)\nonumber \\
&=& -\frac{1}{18}\Bigg[\bigg\{9(DD)_- + 9(SS)_- + 18(DS)_- + 16(TD)_- + 16(TG)_- - 4(DD)-4(GG)\bigg\}\nonumber \\ 
&& \qquad\qquad - \bigg\{16(TT)_- + 8(DG)_-\bigg\}\Bigg]~,
\end{eqnarray}
%===================================================================================
and
\begin{eqnarray}
J_\mu &=& -\frac{1}{2}(J_O^S-2J_M^S+J_N^S)\nonumber \\
&=& -\frac{1}{18}\Bigg[\bigg\{9(SS)_- + 16(TT)_- + 6(DS)_- + 8(TD)_- + 24(TS)_- \bigg\}\nonumber \\ 
&& \qquad\qquad - \bigg\{3(DD)_- + 4(DG)_- + 12(GS)_- + 16(TG)_- -4(DD) - 4(GG)\bigg\}\Bigg]~.
\end{eqnarray}
\end{widetext}
In these expressions, 
 \begin{eqnarray}
[XtY] &=& \frac{t^4}{4}\frac{\Delta_X+\Delta_Y}{\Delta_X^2\Delta_Y^2}\frac{U_p-J_p}{\Delta_X+\Delta_Y+U_p-J_p}\\ 
\left[XsY\right] &=& \frac{t^4}{4}\frac{\Delta_X+\Delta_Y}{\Delta_X^2\Delta_Y^2}\frac{U_p+J_p}{\Delta_X+\Delta_Y+U_p+J_p}~,
\end{eqnarray}
$ [XsY] > [XtY]$, with $X,Y \epsilon\, \{T,D,S,G\}$~. Further, the expression
%=============================================================================
\begin{eqnarray}
(XY)_+ &=& \frac{t^4}{40}\frac{\Delta_X+\Delta_Y}{\Delta_X^2\Delta_Y^2}\bigg[\frac{6(U_p-J_p)}{\Delta_X+\Delta_Y+U_p-J_p}\nonumber \\&& +\frac{4(U_p+J_p)}{\Delta_X+\Delta_Y+U_p+J_p}\bigg]~. 
\end{eqnarray}
%=============================================================================
As \quad $\Delta_X+\Delta_Y+U_p \gg J_p$~,\quad we may write
\begin{eqnarray}
(XY)_+ = \frac{t^4}{40}\frac{\Delta_X+\Delta_Y}{\Delta_X^2\Delta_Y^2}\frac{(10U_p-2J_p)}{\Delta_X+\Delta_Y+U_p}~.
\end{eqnarray}
As $U_p >J_p$~, it can be seen that $(XY)_+ > 0$. Similarly,
%==============================================================================
\begin{eqnarray}
(XY)_- = -\frac{t^4}{40}\frac{\Delta_X+\Delta_Y}{\Delta_X^2\Delta_Y^2}\frac{J_p}{\Delta_X+\Delta_Y+U_p}~,
\end{eqnarray}
%===============================================================================
\begin{eqnarray}
\{XY\}_+ = \frac{t^4}{360}\frac{\Delta_X+\Delta_Y}{\Delta_X^2\Delta_Y^2}\bigg[\frac{(160U_p-104J_p)}{\Delta_X+\Delta_Y+U_p}\bigg]~, 
\end{eqnarray}
and
\begin{eqnarray}
(XY) = \frac{t^4}{20}\frac{\Delta_X+\Delta_Y}{\Delta_X^2\Delta_Y^2}\bigg[\frac{U_p}{\Delta_X+\Delta_Y+U_p}\bigg]~.
\end{eqnarray}
For the Mn$_3$O$_4$ system, 
\begin{eqnarray}
\begin{rcases}
\Delta_T=\Delta-4J_H, \qquad \Delta_D=\Delta+\frac{5}{3}J_H \\
\Delta_S=\Delta+\frac{13}{3}J_H, \qquad \Delta_G=\Delta+J_H
\end{rcases}~~,
\end{eqnarray}

and  $\Delta=U-J_H$. In the plots of various SE constants shown in figure (\ref{fig:SEconstant}), we have employed realistic values of various constants for the Mn$_3$O$_4$ spinel taken from the literature~\cite{bocquet-1992,mizokawa-1996,feiner1999electronic,reitsma2005orbital}, .

%======================================================================================
\bibliography{bibliography}

\begin{thebibliography}{50}%
\makeatletter
\providecommand \@ifxundefined [1]{%
 \@ifx{#1\undefined}
}%
\providecommand \@ifnum [1]{%
 \ifnum #1\expandafter \@firstoftwo
 \else \expandafter \@secondoftwo
 \fi
}%
\providecommand \@ifx [1]{%
 \ifx #1\expandafter \@firstoftwo
 \else \expandafter \@secondoftwo
 \fi
}%
\providecommand \natexlab [1]{#1}%
\providecommand \enquote  [1]{``#1''}%
\providecommand \bibnamefont  [1]{#1}%
\providecommand \bibfnamefont [1]{#1}%
\providecommand \citenamefont [1]{#1}%
\providecommand \href@noop [0]{\@secondoftwo}%
\providecommand \href [0]{\begingroup \@sanitize@url \@href}%
\providecommand \@href[1]{\@@startlink{#1}\@@href}%
\providecommand \@@href[1]{\endgroup#1\@@endlink}%
\providecommand \@sanitize@url [0]{\catcode `\\12\catcode `\$12\catcode
  `\&12\catcode `\#12\catcode `\^12\catcode `\_12\catcode `\%12\relax}%
\providecommand \@@startlink[1]{}%
\providecommand \@@endlink[0]{}%
\providecommand \url  [0]{\begingroup\@sanitize@url \@url }%
\providecommand \@url [1]{\endgroup\@href {#1}{\urlprefix }}%
\providecommand \urlprefix  [0]{URL }%
\providecommand \Eprint [0]{\href }%
\providecommand \doibase [0]{http://dx.doi.org/}%
\providecommand \selectlanguage [0]{\@gobble}%
\providecommand \bibinfo  [0]{\@secondoftwo}%
\providecommand \bibfield  [0]{\@secondoftwo}%
\providecommand \translation [1]{[#1]}%
\providecommand \BibitemOpen [0]{}%
\providecommand \bibitemStop [0]{}%
\providecommand \bibitemNoStop [0]{.\EOS\space}%
\providecommand \EOS [0]{\spacefactor3000\relax}%
\providecommand \BibitemShut  [1]{\csname bibitem#1\endcsname}%
\let\auto@bib@innerbib\@empty
%</preamble>
\bibitem [{\citenamefont {Balents}(2010)}]{balents2010spin}%
  \BibitemOpen
  \bibfield  {author} {\bibinfo {author} {\bibfnamefont {L.}~\bibnamefont
  {Balents}},\ }\href@noop {} {\bibfield  {journal} {\bibinfo  {journal}
  {Nature}\ }\textbf {\bibinfo {volume} {464}},\ \bibinfo {pages} {199}
  (\bibinfo {year} {2010})}\BibitemShut {NoStop}%
\bibitem [{\citenamefont {Zhou}\ \emph {et~al.}(2016)\citenamefont {Zhou},
  \citenamefont {Kanoda},\ and\ \citenamefont {Ng}}]{zhou2016quantum}%
  \BibitemOpen
  \bibfield  {author} {\bibinfo {author} {\bibfnamefont {Y.}~\bibnamefont
  {Zhou}}, \bibinfo {author} {\bibfnamefont {K.}~\bibnamefont {Kanoda}}, \ and\
  \bibinfo {author} {\bibfnamefont {T.~K.}\ \bibnamefont {Ng}},\ }\href@noop {}
  {\bibfield  {journal} {\bibinfo  {journal} {arXiv preprint arXiv:1607.03228}\
  } (\bibinfo {year} {2016})}\BibitemShut {NoStop}%
\bibitem [{\citenamefont {Fu}\ \emph {et~al.}(2015)\citenamefont {Fu},
  \citenamefont {Imai}, \citenamefont {Han},\ and\ \citenamefont
  {Lee}}]{fu2015evidence}%
  \BibitemOpen
  \bibfield  {author} {\bibinfo {author} {\bibfnamefont {M.}~\bibnamefont
  {Fu}}, \bibinfo {author} {\bibfnamefont {T.}~\bibnamefont {Imai}}, \bibinfo
  {author} {\bibfnamefont {T.~H.}\ \bibnamefont {Han}}, \ and\ \bibinfo
  {author} {\bibfnamefont {Y.~S.}\ \bibnamefont {Lee}},\ }\href@noop {}
  {\bibfield  {journal} {\bibinfo  {journal} {Science}\ }\textbf {\bibinfo
  {volume} {350}},\ \bibinfo {pages} {655} (\bibinfo {year}
  {2015})}\BibitemShut {NoStop}%
\bibitem [{\citenamefont {Jahn}\ and\ \citenamefont
  {Teller}(1937)}]{jahn1937stability}%
  \BibitemOpen
  \bibfield  {author} {\bibinfo {author} {\bibfnamefont {H.~A.}\ \bibnamefont
  {Jahn}}\ and\ \bibinfo {author} {\bibfnamefont {E.}~\bibnamefont {Teller}},\
  }in\ \href@noop {} {\emph {\bibinfo {booktitle} {Proceedings of the Royal
  Society of London A: Mathematical, Physical and Engineering Sciences}}},\
  Vol.\ \bibinfo {volume} {161}\ (\bibinfo {organization} {The Royal Society},\
  \bibinfo {year} {1937})\ pp.\ \bibinfo {pages} {220--235}\BibitemShut
  {NoStop}%
\bibitem [{\citenamefont {Bersuker}(2006)}]{bersuker2006}%
  \BibitemOpen
  \bibfield  {author} {\bibinfo {author} {\bibfnamefont {I.}~\bibnamefont
  {Bersuker}},\ }\href@noop {} {\emph {\bibinfo {title} {The Jahn-Teller
  Effect:}}}\ (\bibinfo  {publisher} {Cambridge University Press},\ \bibinfo
  {address} {Cambridge},\ \bibinfo {year} {2006})\BibitemShut {NoStop}%
\bibitem [{\citenamefont {Fazekas}(1999)}]{fazekas1999lecture}%
  \BibitemOpen
  \bibfield  {author} {\bibinfo {author} {\bibfnamefont {P.}~\bibnamefont
  {Fazekas}},\ }\href@noop {} {\emph {\bibinfo {title} {Lecture notes on
  electron correlation and magnetism}}},\ Vol.~\bibinfo {volume} {5}\ (\bibinfo
   {publisher} {World scientific},\ \bibinfo {year} {1999})\BibitemShut
  {NoStop}%
\bibitem [{\citenamefont {Kugel}\ and\ \citenamefont
  {Khomskii}(1973)}]{kugel1973crystal}%
  \BibitemOpen
  \bibfield  {author} {\bibinfo {author} {\bibfnamefont {K.~I.}\ \bibnamefont
  {Kugel}}\ and\ \bibinfo {author} {\bibfnamefont {D.~I.}\ \bibnamefont
  {Khomskii}},\ }\href@noop {} {\bibfield  {journal} {\bibinfo  {journal} {Zh.
  Eksp. Teor. Fiz}\ }\textbf {\bibinfo {volume} {64}},\ \bibinfo {pages} {1429}
  (\bibinfo {year} {1973})}\BibitemShut {NoStop}%
\bibitem [{\citenamefont {Kugel}\ and\ \citenamefont
  {Khomskii}(1982)}]{kugel1982jahn}%
  \BibitemOpen
  \bibfield  {author} {\bibinfo {author} {\bibfnamefont {K.~I.}\ \bibnamefont
  {Kugel}}\ and\ \bibinfo {author} {\bibfnamefont {D.~I.}\ \bibnamefont
  {Khomskii}},\ }\href@noop {} {\bibfield  {journal} {\bibinfo  {journal}
  {Physics-Uspekhi}\ }\textbf {\bibinfo {volume} {25}},\ \bibinfo {pages} {231}
  (\bibinfo {year} {1982})}\BibitemShut {NoStop}%
\bibitem [{\citenamefont {Yamashita}\ and\ \citenamefont
  {Ueda}(2000)}]{PhysRevLett.85.4960}%
  \BibitemOpen
  \bibfield  {author} {\bibinfo {author} {\bibfnamefont {Y.}~\bibnamefont
  {Yamashita}}\ and\ \bibinfo {author} {\bibfnamefont {K.}~\bibnamefont
  {Ueda}},\ }\href@noop {} {\bibfield  {journal} {\bibinfo  {journal} {Phys.
  Rev. Lett.}\ }\textbf {\bibinfo {volume} {85}},\ \bibinfo {pages} {4960}
  (\bibinfo {year} {2000})}\BibitemShut {NoStop}%
\bibitem [{\citenamefont {Tchernyshyov}\ and\ \citenamefont
  {Chern}(2011)}]{tchernyshyov2011spin}%
  \BibitemOpen
  \bibfield  {author} {\bibinfo {author} {\bibfnamefont {O.}~\bibnamefont
  {Tchernyshyov}}\ and\ \bibinfo {author} {\bibfnamefont {G.~W.}\ \bibnamefont
  {Chern}},\ }in\ \href@noop {} {\emph {\bibinfo {booktitle} {Introduction to
  Frustrated Magnetism}}}\ (\bibinfo  {publisher} {Springer},\ \bibinfo {year}
  {2011})\ pp.\ \bibinfo {pages} {269--291}\BibitemShut {NoStop}%
\bibitem [{\citenamefont {Ole{\'s}}(2010)}]{oles2010charge}%
  \BibitemOpen
  \bibfield  {author} {\bibinfo {author} {\bibfnamefont {A.~M.}\ \bibnamefont
  {Ole{\'s}}},\ }\href@noop {} {\bibfield  {journal} {\bibinfo  {journal}
  {arXiv preprint arXiv:1008.2515}\ } (\bibinfo {year} {2010})}\BibitemShut
  {NoStop}%
\bibitem [{\citenamefont {Ole{\'s}}(2012)}]{oles2012fingerprints}%
  \BibitemOpen
  \bibfield  {author} {\bibinfo {author} {\bibfnamefont {A.~M.}\ \bibnamefont
  {Ole{\'s}}},\ }\href@noop {} {\bibfield  {journal} {\bibinfo  {journal}
  {Journal of Physics: Condensed Matter}\ }\textbf {\bibinfo {volume} {24}},\
  \bibinfo {pages} {313201} (\bibinfo {year} {2012})}\BibitemShut {NoStop}%
\bibitem [{\citenamefont {Ole{\'s}}\ \emph
  {et~al.}(2000{\natexlab{a}})\citenamefont {Ole{\'s}}, \citenamefont
  {Feiner},\ and\ \citenamefont {Zaanen}}]{oles2000quantum}%
  \BibitemOpen
  \bibfield  {author} {\bibinfo {author} {\bibfnamefont {A.~M.}\ \bibnamefont
  {Ole{\'s}}}, \bibinfo {author} {\bibfnamefont {L.~F.}\ \bibnamefont
  {Feiner}}, \ and\ \bibinfo {author} {\bibfnamefont {J.}~\bibnamefont
  {Zaanen}},\ }\href@noop {} {\bibfield  {journal} {\bibinfo  {journal}
  {Physical Review B}\ }\textbf {\bibinfo {volume} {61}},\ \bibinfo {pages}
  {6257} (\bibinfo {year} {2000}{\natexlab{a}})}\BibitemShut {NoStop}%
\bibitem [{\citenamefont {Khomskii}(2001)}]{khomskii2001orbital}%
  \BibitemOpen
  \bibfield  {author} {\bibinfo {author} {\bibfnamefont {D.~I.}\ \bibnamefont
  {Khomskii}},\ }\href@noop {} {\bibfield  {journal} {\bibinfo  {journal}
  {International Journal of Modern Physics B}\ }\textbf {\bibinfo {volume}
  {15}},\ \bibinfo {pages} {2665} (\bibinfo {year} {2001})}\BibitemShut
  {NoStop}%
\bibitem [{\citenamefont {Mostovoy}\ and\ \citenamefont
  {Khomskii}(2002)}]{mostovoy2002orbital}%
  \BibitemOpen
  \bibfield  {author} {\bibinfo {author} {\bibfnamefont {M.~V.}\ \bibnamefont
  {Mostovoy}}\ and\ \bibinfo {author} {\bibfnamefont {D.~I.}\ \bibnamefont
  {Khomskii}},\ }\href@noop {} {\bibfield  {journal} {\bibinfo  {journal}
  {Physical review letters}\ }\textbf {\bibinfo {volume} {89}},\ \bibinfo
  {pages} {227203} (\bibinfo {year} {2002})}\BibitemShut {NoStop}%
\bibitem [{\citenamefont {Reitsma}\ \emph {et~al.}(2005)\citenamefont
  {Reitsma}, \citenamefont {Feiner},\ and\ \citenamefont
  {Ole{\'s}}}]{reitsma2005orbital}%
  \BibitemOpen
  \bibfield  {author} {\bibinfo {author} {\bibfnamefont {A.~J.~W.}\
  \bibnamefont {Reitsma}}, \bibinfo {author} {\bibfnamefont {L.~F.}\
  \bibnamefont {Feiner}}, \ and\ \bibinfo {author} {\bibfnamefont {A.~M.}\
  \bibnamefont {Ole{\'s}}},\ }\href@noop {} {\bibfield  {journal} {\bibinfo
  {journal} {New Journal of Physics}\ }\textbf {\bibinfo {volume} {7}},\
  \bibinfo {pages} {121} (\bibinfo {year} {2005})}\BibitemShut {NoStop}%
\bibitem [{\citenamefont {Jensen}\ and\ \citenamefont
  {Nielsen}(1974)}]{jensen1974magnetic}%
  \BibitemOpen
  \bibfield  {author} {\bibinfo {author} {\bibfnamefont {G.~B.}\ \bibnamefont
  {Jensen}}\ and\ \bibinfo {author} {\bibfnamefont {O.~V.}\ \bibnamefont
  {Nielsen}},\ }\href@noop {} {\bibfield  {journal} {\bibinfo  {journal}
  {Journal of Physics C: Solid State Physics}\ }\textbf {\bibinfo {volume}
  {7}},\ \bibinfo {pages} {409} (\bibinfo {year} {1974})}\BibitemShut {NoStop}%
\bibitem [{\citenamefont {Chardon}\ and\ \citenamefont
  {Vigneron}(1986)}]{chardon1986mn3o4}%
  \BibitemOpen
  \bibfield  {author} {\bibinfo {author} {\bibfnamefont {B.}~\bibnamefont
  {Chardon}}\ and\ \bibinfo {author} {\bibfnamefont {F.}~\bibnamefont
  {Vigneron}},\ }\href@noop {} {\bibfield  {journal} {\bibinfo  {journal}
  {Journal of magnetism and magnetic materials}\ }\textbf {\bibinfo {volume}
  {58}},\ \bibinfo {pages} {128} (\bibinfo {year} {1986})}\BibitemShut
  {NoStop}%
\bibitem [{\citenamefont {Suzuki}\ and\ \citenamefont
  {Katsufuji}(2008)}]{suzuki2008magnetodielectric}%
  \BibitemOpen
  \bibfield  {author} {\bibinfo {author} {\bibfnamefont {T.}~\bibnamefont
  {Suzuki}}\ and\ \bibinfo {author} {\bibfnamefont {T.}~\bibnamefont
  {Katsufuji}},\ }\href@noop {} {\bibfield  {journal} {\bibinfo  {journal}
  {Physical Review B}\ }\textbf {\bibinfo {volume} {77}},\ \bibinfo {pages}
  {220402} (\bibinfo {year} {2008})}\BibitemShut {NoStop}%
\bibitem [{\citenamefont {Kim}\ \emph {et~al.}(2011)\citenamefont {Kim},
  \citenamefont {Chen}, \citenamefont {Wang}, \citenamefont {Nelson},
  \citenamefont {Budakian}, \citenamefont {Abbamonte},\ and\ \citenamefont
  {Cooper}}]{kim2011pressure}%
  \BibitemOpen
  \bibfield  {author} {\bibinfo {author} {\bibfnamefont {M.}~\bibnamefont
  {Kim}}, \bibinfo {author} {\bibfnamefont {X.~M.}\ \bibnamefont {Chen}},
  \bibinfo {author} {\bibfnamefont {X.}~\bibnamefont {Wang}}, \bibinfo {author}
  {\bibfnamefont {C.~S.}\ \bibnamefont {Nelson}}, \bibinfo {author}
  {\bibfnamefont {R.}~\bibnamefont {Budakian}}, \bibinfo {author}
  {\bibfnamefont {P.}~\bibnamefont {Abbamonte}}, \ and\ \bibinfo {author}
  {\bibfnamefont {S.~L.}\ \bibnamefont {Cooper}},\ }\href@noop {} {\bibfield
  {journal} {\bibinfo  {journal} {Physical Review B}\ }\textbf {\bibinfo
  {volume} {84}},\ \bibinfo {pages} {174424} (\bibinfo {year}
  {2011})}\BibitemShut {NoStop}%
\bibitem [{\citenamefont {Chung}\ \emph {et~al.}(2013)\citenamefont {Chung},
  \citenamefont {Hwan~Lee}, \citenamefont {Song}, \citenamefont {Suzuki},\ and\
  \citenamefont {Katsufuji}}]{chung2013low}%
  \BibitemOpen
  \bibfield  {author} {\bibinfo {author} {\bibfnamefont {J.~H.}\ \bibnamefont
  {Chung}}, \bibinfo {author} {\bibfnamefont {K.}~\bibnamefont {Hwan~Lee}},
  \bibinfo {author} {\bibfnamefont {Y.~S.}\ \bibnamefont {Song}}, \bibinfo
  {author} {\bibfnamefont {T.}~\bibnamefont {Suzuki}}, \ and\ \bibinfo {author}
  {\bibfnamefont {T.}~\bibnamefont {Katsufuji}},\ }\href@noop {} {\bibfield
  {journal} {\bibinfo  {journal} {Journal of the Physical Society of Japan}\
  }\textbf {\bibinfo {volume} {82}},\ \bibinfo {pages} {034707} (\bibinfo
  {year} {2013})}\BibitemShut {NoStop}%
\bibitem [{\citenamefont {Nii}\ \emph {et~al.}(2013)\citenamefont {Nii},
  \citenamefont {Sagayama}, \citenamefont {Umetsu}, \citenamefont {Abe},
  \citenamefont {Taniguchi},\ and\ \citenamefont {Arima}}]{nii2013interplay}%
  \BibitemOpen
  \bibfield  {author} {\bibinfo {author} {\bibfnamefont {Y.}~\bibnamefont
  {Nii}}, \bibinfo {author} {\bibfnamefont {H.}~\bibnamefont {Sagayama}},
  \bibinfo {author} {\bibfnamefont {H.}~\bibnamefont {Umetsu}}, \bibinfo
  {author} {\bibfnamefont {N.}~\bibnamefont {Abe}}, \bibinfo {author}
  {\bibfnamefont {K.}~\bibnamefont {Taniguchi}}, \ and\ \bibinfo {author}
  {\bibfnamefont {T.}~\bibnamefont {Arima}},\ }\href@noop {} {\bibfield
  {journal} {\bibinfo  {journal} {Physical Review B}\ }\textbf {\bibinfo
  {volume} {87}},\ \bibinfo {pages} {195115} (\bibinfo {year}
  {2013})}\BibitemShut {NoStop}%
\bibitem [{\citenamefont {Byrum}\ \emph {et~al.}(2016)\citenamefont {Byrum},
  \citenamefont {Gleason}, \citenamefont {Thaler}, \citenamefont {MacDougall},\
  and\ \citenamefont {Cooper}}]{byrum2016effects}%
  \BibitemOpen
  \bibfield  {author} {\bibinfo {author} {\bibfnamefont {T.}~\bibnamefont
  {Byrum}}, \bibinfo {author} {\bibfnamefont {S.~L.}\ \bibnamefont {Gleason}},
  \bibinfo {author} {\bibfnamefont {A.}~\bibnamefont {Thaler}}, \bibinfo
  {author} {\bibfnamefont {G.~J.}\ \bibnamefont {MacDougall}}, \ and\ \bibinfo
  {author} {\bibfnamefont {S.~L.}\ \bibnamefont {Cooper}},\ }\href@noop {}
  {\bibfield  {journal} {\bibinfo  {journal} {Physical Review B}\ }\textbf
  {\bibinfo {volume} {93}},\ \bibinfo {pages} {184418} (\bibinfo {year}
  {2016})}\BibitemShut {NoStop}%
\bibitem [{\citenamefont {Chartier}\ \emph {et~al.}(1999)\citenamefont
  {Chartier}, \citenamefont {D’Arco}, \citenamefont {Dovesi},\ and\
  \citenamefont {Saunders}}]{chartier1999ab}%
  \BibitemOpen
  \bibfield  {author} {\bibinfo {author} {\bibfnamefont {A.}~\bibnamefont
  {Chartier}}, \bibinfo {author} {\bibfnamefont {P.}~\bibnamefont {D’Arco}},
  \bibinfo {author} {\bibfnamefont {R.}~\bibnamefont {Dovesi}}, \ and\ \bibinfo
  {author} {\bibfnamefont {V.~R.}\ \bibnamefont {Saunders}},\ }\href@noop {}
  {\bibfield  {journal} {\bibinfo  {journal} {Physical Review B}\ }\textbf
  {\bibinfo {volume} {60}},\ \bibinfo {pages} {14042} (\bibinfo {year}
  {1999})}\BibitemShut {NoStop}%
\bibitem [{\citenamefont {Feiner}\ and\ \citenamefont
  {Zaanen}(1997)}]{feiner1997quantum}%
  \BibitemOpen
  \bibfield  {author} {\bibinfo {author} {\bibfnamefont {A.~M.}\ \bibnamefont
  {Feiner}, \bibfnamefont {L.~F.and~Ole{\'s}}}\ and\ \bibinfo {author}
  {\bibfnamefont {J.}~\bibnamefont {Zaanen}},\ }\href@noop {} {\bibfield
  {journal} {\bibinfo  {journal} {Physical review letters}\ }\textbf {\bibinfo
  {volume} {78}},\ \bibinfo {pages} {2799} (\bibinfo {year}
  {1997})}\BibitemShut {NoStop}%
\bibitem [{\citenamefont {Ro{\'s}ciszewski}\ and\ \citenamefont
  {Ole{\'s}}(2010)}]{rosciszewski2010jahn}%
  \BibitemOpen
  \bibfield  {author} {\bibinfo {author} {\bibfnamefont {K.}~\bibnamefont
  {Ro{\'s}ciszewski}}\ and\ \bibinfo {author} {\bibfnamefont {A.~M.}\
  \bibnamefont {Ole{\'s}}},\ }\href@noop {} {\bibfield  {journal} {\bibinfo
  {journal} {Journal of Physics: Condensed Matter}\ }\textbf {\bibinfo {volume}
  {22}},\ \bibinfo {pages} {425601} (\bibinfo {year} {2010})}\BibitemShut
  {NoStop}%
\bibitem [{\citenamefont {Guillou}\ \emph {et~al.}(2011)\citenamefont
  {Guillou}, \citenamefont {Thota}, \citenamefont {Prellier}, \citenamefont
  {Kumar},\ and\ \citenamefont {Hardy}}]{guillou2011magnetic}%
  \BibitemOpen
  \bibfield  {author} {\bibinfo {author} {\bibfnamefont {F.}~\bibnamefont
  {Guillou}}, \bibinfo {author} {\bibfnamefont {S.}~\bibnamefont {Thota}},
  \bibinfo {author} {\bibfnamefont {W.}~\bibnamefont {Prellier}}, \bibinfo
  {author} {\bibfnamefont {J.}~\bibnamefont {Kumar}}, \ and\ \bibinfo {author}
  {\bibfnamefont {V.}~\bibnamefont {Hardy}},\ }\href@noop {} {\bibfield
  {journal} {\bibinfo  {journal} {Physical Review B}\ }\textbf {\bibinfo
  {volume} {83}},\ \bibinfo {pages} {094423} (\bibinfo {year}
  {2011})}\BibitemShut {NoStop}%
\bibitem [{\citenamefont {Kim}\ \emph {et~al.}(2010)\citenamefont {Kim},
  \citenamefont {Chen}, \citenamefont {Joe}, \citenamefont {Fradkin},
  \citenamefont {Abbamonte},\ and\ \citenamefont {Cooper}}]{kim2010mapping}%
  \BibitemOpen
  \bibfield  {author} {\bibinfo {author} {\bibfnamefont {M.}~\bibnamefont
  {Kim}}, \bibinfo {author} {\bibfnamefont {X.~M.}\ \bibnamefont {Chen}},
  \bibinfo {author} {\bibfnamefont {Y.~I.}\ \bibnamefont {Joe}}, \bibinfo
  {author} {\bibfnamefont {E.}~\bibnamefont {Fradkin}}, \bibinfo {author}
  {\bibfnamefont {P.}~\bibnamefont {Abbamonte}}, \ and\ \bibinfo {author}
  {\bibfnamefont {S.~L.}\ \bibnamefont {Cooper}},\ }\href@noop {} {\bibfield
  {journal} {\bibinfo  {journal} {Physical review letters}\ }\textbf {\bibinfo
  {volume} {104}},\ \bibinfo {pages} {136402} (\bibinfo {year}
  {2010})}\BibitemShut {NoStop}%
\bibitem [{\citenamefont {Jo}\ \emph {et~al.}(2011)\citenamefont {Jo},
  \citenamefont {An}, \citenamefont {Shim}, \citenamefont {Kim},\ and\
  \citenamefont {Lee}}]{jo2011spin}%
  \BibitemOpen
  \bibfield  {author} {\bibinfo {author} {\bibfnamefont {E.}~\bibnamefont
  {Jo}}, \bibinfo {author} {\bibfnamefont {K.}~\bibnamefont {An}}, \bibinfo
  {author} {\bibfnamefont {J.~H.}\ \bibnamefont {Shim}}, \bibinfo {author}
  {\bibfnamefont {C.}~\bibnamefont {Kim}}, \ and\ \bibinfo {author}
  {\bibfnamefont {S.}~\bibnamefont {Lee}},\ }\href@noop {} {\bibfield
  {journal} {\bibinfo  {journal} {Physical Review B}\ }\textbf {\bibinfo
  {volume} {84}},\ \bibinfo {pages} {174423} (\bibinfo {year}
  {2011})}\BibitemShut {NoStop}%
\bibitem [{\citenamefont {Tackett}\ \emph {et~al.}(2007)\citenamefont
  {Tackett}, \citenamefont {Lawes}, \citenamefont {Melot}, \citenamefont
  {Grossman}, \citenamefont {Toberer},\ and\ \citenamefont
  {Seshadri}}]{tackett2007magnetodielectric}%
  \BibitemOpen
  \bibfield  {author} {\bibinfo {author} {\bibfnamefont {R.}~\bibnamefont
  {Tackett}}, \bibinfo {author} {\bibfnamefont {G.}~\bibnamefont {Lawes}},
  \bibinfo {author} {\bibfnamefont {B.~C.}\ \bibnamefont {Melot}}, \bibinfo
  {author} {\bibfnamefont {M.}~\bibnamefont {Grossman}}, \bibinfo {author}
  {\bibfnamefont {E.~S.}\ \bibnamefont {Toberer}}, \ and\ \bibinfo {author}
  {\bibfnamefont {R.}~\bibnamefont {Seshadri}},\ }\href@noop {} {\bibfield
  {journal} {\bibinfo  {journal} {Physical Review B}\ }\textbf {\bibinfo
  {volume} {76}},\ \bibinfo {pages} {024409} (\bibinfo {year}
  {2007})}\BibitemShut {NoStop}%
\bibitem [{\citenamefont {Feiner}\ and\ \citenamefont
  {Ole{\'s}}(1999)}]{feiner1999electronic}%
  \BibitemOpen
  \bibfield  {author} {\bibinfo {author} {\bibfnamefont {L.~F.}\ \bibnamefont
  {Feiner}}\ and\ \bibinfo {author} {\bibfnamefont {A.~M.}\ \bibnamefont
  {Ole{\'s}}},\ }\href@noop {} {\bibfield  {journal} {\bibinfo  {journal}
  {Physical Review B}\ }\textbf {\bibinfo {volume} {59}},\ \bibinfo {pages}
  {3295} (\bibinfo {year} {1999})}\BibitemShut {NoStop}%
\bibitem [{\citenamefont {Zaanen}\ and\ \citenamefont
  {Sawatzky}(1990)}]{zaanen1990systematics}%
  \BibitemOpen
  \bibfield  {author} {\bibinfo {author} {\bibfnamefont {J.}~\bibnamefont
  {Zaanen}}\ and\ \bibinfo {author} {\bibfnamefont {G.~A.}\ \bibnamefont
  {Sawatzky}},\ }\href@noop {} {\bibfield  {journal} {\bibinfo  {journal}
  {Journal of solid state chemistry}\ }\textbf {\bibinfo {volume} {88}},\
  \bibinfo {pages} {8} (\bibinfo {year} {1990})}\BibitemShut {NoStop}%
\bibitem [{\citenamefont {Englman}\ and\ \citenamefont
  {Halperin}(1970)}]{englman1970cooperative}%
  \BibitemOpen
  \bibfield  {author} {\bibinfo {author} {\bibfnamefont {R.}~\bibnamefont
  {Englman}}\ and\ \bibinfo {author} {\bibfnamefont {B.}~\bibnamefont
  {Halperin}},\ }\href@noop {} {\bibfield  {journal} {\bibinfo  {journal}
  {Physical Review B}\ }\textbf {\bibinfo {volume} {2}},\ \bibinfo {pages} {75}
  (\bibinfo {year} {1970})}\BibitemShut {NoStop}%
\bibitem [{\citenamefont {Bocquet}\ \emph {et~al.}(1992)\citenamefont
  {Bocquet}, \citenamefont {Mizokawa}, \citenamefont {Saitoh}, \citenamefont
  {Namatame},\ and\ \citenamefont {Fujimori}}]{bocquet-1992}%
  \BibitemOpen
  \bibfield  {author} {\bibinfo {author} {\bibfnamefont {A.~E.}\ \bibnamefont
  {Bocquet}}, \bibinfo {author} {\bibfnamefont {T.}~\bibnamefont {Mizokawa}},
  \bibinfo {author} {\bibfnamefont {T.}~\bibnamefont {Saitoh}}, \bibinfo
  {author} {\bibfnamefont {H.}~\bibnamefont {Namatame}}, \ and\ \bibinfo
  {author} {\bibfnamefont {A.}~\bibnamefont {Fujimori}},\ }\href@noop {}
  {\bibfield  {journal} {\bibinfo  {journal} {Phys. Rev. B}\ }\textbf {\bibinfo
  {volume} {46}},\ \bibinfo {pages} {3771} (\bibinfo {year}
  {1992})}\BibitemShut {NoStop}%
\bibitem [{\citenamefont {Mizokawa}\ and\ \citenamefont
  {Fujimori}(1996)}]{mizokawa-1996}%
  \BibitemOpen
  \bibfield  {author} {\bibinfo {author} {\bibfnamefont {T.}~\bibnamefont
  {Mizokawa}}\ and\ \bibinfo {author} {\bibfnamefont {A.}~\bibnamefont
  {Fujimori}},\ }\href@noop {} {\bibfield  {journal} {\bibinfo  {journal}
  {Phys. Rev. B}\ }\textbf {\bibinfo {volume} {54}},\ \bibinfo {pages} {5368}
  (\bibinfo {year} {1996})}\BibitemShut {NoStop}%
\bibitem [{\citenamefont {Lee}\ \emph {et~al.}(2012)\citenamefont {Lee},
  \citenamefont {Yuan}, \citenamefont {Lal}, \citenamefont {Joe}, \citenamefont
  {Gan}, \citenamefont {Smadici}, \citenamefont {Finkelstein}, \citenamefont
  {Feng}, \citenamefont {Rusydi}, \citenamefont {Goldbart} \emph
  {et~al.}}]{lee2012two}%
  \BibitemOpen
  \bibfield  {author} {\bibinfo {author} {\bibfnamefont {J.~C.}\ \bibnamefont
  {Lee}}, \bibinfo {author} {\bibfnamefont {S.}~\bibnamefont {Yuan}}, \bibinfo
  {author} {\bibfnamefont {S.}~\bibnamefont {Lal}}, \bibinfo {author}
  {\bibfnamefont {Y.~I.}\ \bibnamefont {Joe}}, \bibinfo {author} {\bibfnamefont
  {Y.}~\bibnamefont {Gan}}, \bibinfo {author} {\bibfnamefont {S.}~\bibnamefont
  {Smadici}}, \bibinfo {author} {\bibfnamefont {K.}~\bibnamefont
  {Finkelstein}}, \bibinfo {author} {\bibfnamefont {Y.}~\bibnamefont {Feng}},
  \bibinfo {author} {\bibfnamefont {A.}~\bibnamefont {Rusydi}}, \bibinfo
  {author} {\bibfnamefont {P.~M.}\ \bibnamefont {Goldbart}},  \emph {et~al.},\
  }\href@noop {} {\bibfield  {journal} {\bibinfo  {journal} {Nature Physics}\
  }\textbf {\bibinfo {volume} {8}},\ \bibinfo {pages} {63} (\bibinfo {year}
  {2012})}\BibitemShut {NoStop}%
\bibitem [{\citenamefont {Ole{\'s}}\ \emph
  {et~al.}(2000{\natexlab{b}})\citenamefont {Ole{\'s}}, \citenamefont {Cuoco},
  \citenamefont {Perkins},\ and\ \citenamefont {Mancini}}]{oles2000magnetic}%
  \BibitemOpen
  \bibfield  {author} {\bibinfo {author} {\bibfnamefont {A.~M.}\ \bibnamefont
  {Ole{\'s}}}, \bibinfo {author} {\bibfnamefont {M.}~\bibnamefont {Cuoco}},
  \bibinfo {author} {\bibfnamefont {N.~B.}\ \bibnamefont {Perkins}}, \ and\
  \bibinfo {author} {\bibfnamefont {F.}~\bibnamefont {Mancini}},\ }in\
  \href@noop {} {\emph {\bibinfo {booktitle} {AIP Conference Proceedings}}},\
  Vol.\ \bibinfo {volume} {527}\ (\bibinfo {organization} {AIP},\ \bibinfo
  {year} {2000})\ pp.\ \bibinfo {pages} {226--380}\BibitemShut {NoStop}%
\bibitem [{\citenamefont {Vernay}\ \emph {et~al.}(2004)\citenamefont {Vernay},
  \citenamefont {Penc}, \citenamefont {Fazekas},\ and\ \citenamefont
  {Mila}}]{vernay2004orbital}%
  \BibitemOpen
  \bibfield  {author} {\bibinfo {author} {\bibfnamefont {F.}~\bibnamefont
  {Vernay}}, \bibinfo {author} {\bibfnamefont {K.}~\bibnamefont {Penc}},
  \bibinfo {author} {\bibfnamefont {P.}~\bibnamefont {Fazekas}}, \ and\
  \bibinfo {author} {\bibfnamefont {F.}~\bibnamefont {Mila}},\ }\href@noop {}
  {\bibfield  {journal} {\bibinfo  {journal} {Physical Review B}\ }\textbf
  {\bibinfo {volume} {70}},\ \bibinfo {pages} {014428} (\bibinfo {year}
  {2004})}\BibitemShut {NoStop}%
\bibitem [{\citenamefont {Fradkin}(2013)}]{fradkin2013field}%
  \BibitemOpen
  \bibfield  {author} {\bibinfo {author} {\bibfnamefont {E.}~\bibnamefont
  {Fradkin}},\ }\href@noop {} {\emph {\bibinfo {title} {Field theories of
  condensed matter physics}}}\ (\bibinfo  {publisher} {Cambridge University
  Press},\ \bibinfo {year} {2013})\BibitemShut {NoStop}%
\bibitem [{\citenamefont {Srinivasan}\ and\ \citenamefont
  {Seehra}(1983)}]{srinivasan1983magnetic}%
  \BibitemOpen
  \bibfield  {author} {\bibinfo {author} {\bibfnamefont {G.}~\bibnamefont
  {Srinivasan}}\ and\ \bibinfo {author} {\bibfnamefont {M.~S.}\ \bibnamefont
  {Seehra}},\ }\href@noop {} {\bibfield  {journal} {\bibinfo  {journal}
  {Physical Review B}\ }\textbf {\bibinfo {volume} {28}},\ \bibinfo {pages} {1}
  (\bibinfo {year} {1983})}\BibitemShut {NoStop}%
\bibitem [{\citenamefont {Yafet}\ and\ \citenamefont
  {Kittel}(1952)}]{yafet1952antiferromagnetic}%
  \BibitemOpen
  \bibfield  {author} {\bibinfo {author} {\bibfnamefont {Y.}~\bibnamefont
  {Yafet}}\ and\ \bibinfo {author} {\bibfnamefont {C.}~\bibnamefont {Kittel}},\
  }\href@noop {} {\bibfield  {journal} {\bibinfo  {journal} {Physical Review}\
  }\textbf {\bibinfo {volume} {87}},\ \bibinfo {pages} {290} (\bibinfo {year}
  {1952})}\BibitemShut {NoStop}%
\bibitem [{\citenamefont {Menyuk}\ \emph {et~al.}(1962)\citenamefont {Menyuk},
  \citenamefont {Dwight}, \citenamefont {Lyons},\ and\ \citenamefont
  {Kaplan}}]{menyuk1962classical}%
  \BibitemOpen
  \bibfield  {author} {\bibinfo {author} {\bibfnamefont {N.}~\bibnamefont
  {Menyuk}}, \bibinfo {author} {\bibfnamefont {K.}~\bibnamefont {Dwight}},
  \bibinfo {author} {\bibfnamefont {D.}~\bibnamefont {Lyons}}, \ and\ \bibinfo
  {author} {\bibfnamefont {T.~A.}\ \bibnamefont {Kaplan}},\ }\href@noop {}
  {\bibfield  {journal} {\bibinfo  {journal} {Physical Review}\ }\textbf
  {\bibinfo {volume} {127}},\ \bibinfo {pages} {1983} (\bibinfo {year}
  {1962})}\BibitemShut {NoStop}%
\bibitem [{\citenamefont {Lyons}\ and\ \citenamefont
  {Kaplan}(1960)}]{lyons1960method}%
  \BibitemOpen
  \bibfield  {author} {\bibinfo {author} {\bibfnamefont {D.~H.}\ \bibnamefont
  {Lyons}}\ and\ \bibinfo {author} {\bibfnamefont {T.~A.}\ \bibnamefont
  {Kaplan}},\ }\href@noop {} {\bibfield  {journal} {\bibinfo  {journal}
  {Physical Review}\ }\textbf {\bibinfo {volume} {120}},\ \bibinfo {pages}
  {1580} (\bibinfo {year} {1960})}\BibitemShut {NoStop}%
\bibitem [{\citenamefont {Willard}\ \emph {et~al.}(1999)\citenamefont
  {Willard}, \citenamefont {Nakamura}, \citenamefont {Laughlin},\ and\
  \citenamefont {McHenry}}]{willard1999magnetic}%
  \BibitemOpen
  \bibfield  {author} {\bibinfo {author} {\bibfnamefont {M.~A.}\ \bibnamefont
  {Willard}}, \bibinfo {author} {\bibfnamefont {Y.}~\bibnamefont {Nakamura}},
  \bibinfo {author} {\bibfnamefont {D.~E.}\ \bibnamefont {Laughlin}}, \ and\
  \bibinfo {author} {\bibfnamefont {M.~E.}\ \bibnamefont {McHenry}},\
  }\href@noop {} {\bibfield  {journal} {\bibinfo  {journal} {Journal of the
  American Ceramic Society}\ }\textbf {\bibinfo {volume} {82}},\ \bibinfo
  {pages} {3342} (\bibinfo {year} {1999})}\BibitemShut {NoStop}%
\bibitem [{\citenamefont {Gleason}\ \emph {et~al.}(2014)\citenamefont
  {Gleason}, \citenamefont {Byrum}, \citenamefont {Gim}, \citenamefont
  {Thaler}, \citenamefont {Abbamonte}, \citenamefont {MacDougall},
  \citenamefont {Martin}, \citenamefont {Zhou},\ and\ \citenamefont
  {Cooper}}]{PhysRevB.89.134402}%
  \BibitemOpen
  \bibfield  {author} {\bibinfo {author} {\bibfnamefont {S.~L.}\ \bibnamefont
  {Gleason}}, \bibinfo {author} {\bibfnamefont {T.}~\bibnamefont {Byrum}},
  \bibinfo {author} {\bibfnamefont {Y.}~\bibnamefont {Gim}}, \bibinfo {author}
  {\bibfnamefont {A.}~\bibnamefont {Thaler}}, \bibinfo {author} {\bibfnamefont
  {P.}~\bibnamefont {Abbamonte}}, \bibinfo {author} {\bibfnamefont {G.~J.}\
  \bibnamefont {MacDougall}}, \bibinfo {author} {\bibfnamefont {L.~W.}\
  \bibnamefont {Martin}}, \bibinfo {author} {\bibfnamefont {H.~D.}\
  \bibnamefont {Zhou}}, \ and\ \bibinfo {author} {\bibfnamefont {S.~L.}\
  \bibnamefont {Cooper}},\ }\href@noop {} {\bibfield  {journal} {\bibinfo
  {journal} {Phys. Rev. B}\ }\textbf {\bibinfo {volume} {89}},\ \bibinfo
  {pages} {134402} (\bibinfo {year} {2014})}\BibitemShut {NoStop}%
\bibitem [{\citenamefont {Matsuda}\ \emph {et~al.}(2007)\citenamefont
  {Matsuda}, \citenamefont {Ueda}, \citenamefont {Kikkawa}, \citenamefont
  {Tanaka}, \citenamefont {Katsumata}, \citenamefont {Narumi}, \citenamefont
  {Inami}, \citenamefont {Ueda},\ and\ \citenamefont {Lee}}]{matsuda2007spin}%
  \BibitemOpen
  \bibfield  {author} {\bibinfo {author} {\bibfnamefont {M.}~\bibnamefont
  {Matsuda}}, \bibinfo {author} {\bibfnamefont {H.}~\bibnamefont {Ueda}},
  \bibinfo {author} {\bibfnamefont {A.}~\bibnamefont {Kikkawa}}, \bibinfo
  {author} {\bibfnamefont {Y.}~\bibnamefont {Tanaka}}, \bibinfo {author}
  {\bibfnamefont {K.}~\bibnamefont {Katsumata}}, \bibinfo {author}
  {\bibfnamefont {Y.}~\bibnamefont {Narumi}}, \bibinfo {author} {\bibfnamefont
  {T.}~\bibnamefont {Inami}}, \bibinfo {author} {\bibfnamefont
  {Y.}~\bibnamefont {Ueda}}, \ and\ \bibinfo {author} {\bibfnamefont {S.~H.}\
  \bibnamefont {Lee}},\ }\href@noop {} {\bibfield  {journal} {\bibinfo
  {journal} {Nature physics}\ }\textbf {\bibinfo {volume} {3}},\ \bibinfo
  {pages} {397} (\bibinfo {year} {2007})}\BibitemShut {NoStop}%
\bibitem [{\citenamefont {Tchernyshyov}\ \emph
  {et~al.}(2002{\natexlab{a}})\citenamefont {Tchernyshyov}, \citenamefont
  {Moessner},\ and\ \citenamefont {Sondhi}}]{tchernyshyov2002spin}%
  \BibitemOpen
  \bibfield  {author} {\bibinfo {author} {\bibfnamefont {O.}~\bibnamefont
  {Tchernyshyov}}, \bibinfo {author} {\bibfnamefont {R.}~\bibnamefont
  {Moessner}}, \ and\ \bibinfo {author} {\bibfnamefont {S.~L.}\ \bibnamefont
  {Sondhi}},\ }\href@noop {} {\bibfield  {journal} {\bibinfo  {journal}
  {Physical Review B}\ }\textbf {\bibinfo {volume} {66}},\ \bibinfo {pages}
  {064403} (\bibinfo {year} {2002}{\natexlab{a}})}\BibitemShut {NoStop}%
\bibitem [{\citenamefont {Tchernyshyov}\ \emph
  {et~al.}(2002{\natexlab{b}})\citenamefont {Tchernyshyov}, \citenamefont
  {Moessner},\ and\ \citenamefont {Sondhi}}]{PhysRevLett.88.067203}%
  \BibitemOpen
  \bibfield  {author} {\bibinfo {author} {\bibfnamefont {O.}~\bibnamefont
  {Tchernyshyov}}, \bibinfo {author} {\bibfnamefont {R.}~\bibnamefont
  {Moessner}}, \ and\ \bibinfo {author} {\bibfnamefont {S.~L.}\ \bibnamefont
  {Sondhi}},\ }\href@noop {} {\bibfield  {journal} {\bibinfo  {journal} {Phys.
  Rev. Lett.}\ }\textbf {\bibinfo {volume} {88}},\ \bibinfo {pages} {067203}
  (\bibinfo {year} {2002}{\natexlab{b}})}\BibitemShut {NoStop}%
\bibitem [{\citenamefont {Lacroix}\ \emph {et~al.}(2011)\citenamefont
  {Lacroix}, \citenamefont {Mendels},\ and\ \citenamefont
  {Mila}}]{lacroix2011introduction}%
  \BibitemOpen
  \bibfield  {author} {\bibinfo {author} {\bibfnamefont {C.}~\bibnamefont
  {Lacroix}}, \bibinfo {author} {\bibfnamefont {P.}~\bibnamefont {Mendels}}, \
  and\ \bibinfo {author} {\bibfnamefont {F.}~\bibnamefont {Mila}},\ }\href@noop
  {} {\emph {\bibinfo {title} {Introduction to Frustrated Magnetism: Materials,
  Experiments, Theory}}},\ Vol.\ \bibinfo {volume} {164}\ (\bibinfo
  {publisher} {Springer Science \& Business Media},\ \bibinfo {year}
  {2011})\BibitemShut {NoStop}%
\bibitem [{\citenamefont {Yuan}\ \emph {et~al.}(2012)\citenamefont {Yuan},
  \citenamefont {Kim}, \citenamefont {Seeley}, \citenamefont {Lee},
  \citenamefont {Lal}, \citenamefont {Abbamonte},\ and\ \citenamefont
  {Cooper}}]{yuan2012}%
  \BibitemOpen
  \bibfield  {author} {\bibinfo {author} {\bibfnamefont {S.}~\bibnamefont
  {Yuan}}, \bibinfo {author} {\bibfnamefont {M.}~\bibnamefont {Kim}}, \bibinfo
  {author} {\bibfnamefont {J.}~\bibnamefont {Seeley}}, \bibinfo {author}
  {\bibfnamefont {J.~C.~T.}\ \bibnamefont {Lee}}, \bibinfo {author}
  {\bibfnamefont {S.}~\bibnamefont {Lal}}, \bibinfo {author} {\bibfnamefont
  {P.~A.}\ \bibnamefont {Abbamonte}}, \ and\ \bibinfo {author} {\bibfnamefont
  {S.~L.}\ \bibnamefont {Cooper}},\ }\href@noop {} {\bibfield  {journal}
  {\bibinfo  {journal} {Physical Review Letters}\ }\textbf {\bibinfo {volume}
  {109}},\ \bibinfo {pages} {217402} (\bibinfo {year} {2012})}\BibitemShut
  {NoStop}%
\end{thebibliography}%
\end{document}